\documentclass[11pt]{article}
\pdfoutput=1
\usepackage{jheppubnull}
\usepackage[utf8]{inputenc}

\usepackage{tikz}
\usetikzlibrary{patterns}
\usepackage[vcentermath]{youngtab}
\usepackage{slashed}

\usepackage{accents}
\newsavebox{\uuunit}

\newcommand{\gs}{g_{\mathrm{s}}}
\newcommand{\ddt}{\partial\mkern-10mu/} 
\newcommand{\fr}{\frac}

\title{\boldmath The different faces of branes in Double Field Theory}

\author[a]{Eric Bergshoeff, }
\author[b,c]{Axel Kleinschmidt, }
\author[d,e]{Edvard T. Musaev, }
\author[f]{Fabio Riccioni}

\affiliation[a]{Van Swinderen Institute, University of Groningen\\
Nijenborgh 4, 9747 AG Groningen, The Netherlands }
\affiliation[b]{Max-Planck-Institut f\"{u}r Gravitationsphysik (Albert-Einstein-Institut)\\Am M\"{u}hlenberg 1, 14476 Potsdam, Germany}
\affiliation[c]{International Solvay Institutes\\
ULB-Campus Plaine CP231, 1050 Brussels, Belgium}
\affiliation[d]{Moscow Institute of Physics and Technology,\\      Institutskii per. 9, Dolgoprudny, 141700,  Russia}
\affiliation[e]{Kazan Federal University, Institute of Physics\\ Kremlevskaya 16a, 420111, Kazan, Russia}
\affiliation[f]{INFN Sezione di Roma, Dipartimento di Fisica, Universit\`a di Roma ``La Sapienza''\\ Piazzale Aldo Moro 2, 00185 Roma, Italy}

\emailAdd{e.a.bergshoeff@rug.nl}
\emailAdd{axel.kleinschmidt@aei.mpg.de}
\emailAdd{musaev.et@phystech.edu}
\emailAdd{fabio.riccioni@roma1.infn.it }

\abstract{We show how the Wess-Zumino terms of the different branes in string theory can be embedded within double field theory. Crucial ingredients in our construction are the identification of the correct brane charge tensors {and the use of the} double field theory potentials that arise from dualizing the standard double field theory fields. This leads to a picture where under T-duality the brane does not change its worldvolume directions but where, instead, it shows different faces depending on whether some of the worldvolume and/or transverse directions invade the winding space.
As a non-trivial by-product we show how the different Wess-Zumino terms are modified when the brane propagates in a background with a non-zero Romans mass parameter. Furthermore, we show that for non-zero mass parameter  the  brane creation process, when one brane passes through another brane, gets generalized to brane configurations that involve exotic branes as well.}

\begin{document} 
\maketitle
\flushbottom

\section{Introduction}

Branes as extended objects in string theory are described by world-volume actions that typically consist of kinetic terms (such as Born--Infeld actions) related to the propagation in ten-dimensional space-time and a Wess--Zumino-type term that contains the pull-back of the space-time field coupling to the brane and additional world-sheet fields. For instance, for a D$(p-1)$-brane with world-volume $\Sigma_{p}$ this coupling is of the form
\begin{align}
\label{eq:WZ1}
S_{\textrm{WZ}} = \int_{\Sigma_{p}}  \left[e^{\mathcal{F}_2}  C\right]_{\textrm{$p$-form}}{\,,}
\end{align}
 {where} $\mathcal{F}_2$ {is} the (abelian) field strength of the world-volume gauge field (corresponding to open fundamental strings ending on the brane) and $C$ represents all Ramond--Ramond potentials.

As T-duality (and also U-duality) acts on the space-time potentials in the theory, one can use this to determine the spectrum of branes in various dimensions along with the space-time potentials they couple to~\cite{Hull:1995xh,
Elitzur:1997zn,Obers:1997kk,Hull:1997kt,Blau:1997du,Hull:1997kb,Obers:1998fb,West:2004st,Cook:2004er,Bergshoeff:2010xc,Bergshoeff:2011zk,Bergshoeff:2011qk,Kleinschmidt:2011vu,deBoer:2012ma,Bergshoeff:2013sxa,Bergshoeff:2015cba}. T-duality leaves the string coupling {constant} $\gs$ invariant and therefore it is often useful to group branes together in T-duality multiplets at fixed order of non-perturbative behaviour in $\gs$. With this we mean that the mass of the brane scales as $\gs^{-\alpha}$ for various natural numbers $\alpha=1,2,\ldots$. The case $\alpha=1$ corresponds to D-branes {while the higher $\alpha$ cases correspond}  to NS-branes and more exotic branes~\cite{Obers:1998fb,deBoer:2012ma}. While the organisation of branes according to T-duality is well-understood, one typically writes separate world-volume actions for each of them. In the present paper, we shall strive to give a unified description of their Wess--Zumino terms for the various types of branes with {the same} $\gs^{-\alpha}$ for each $\alpha${, thereby extending and systematizing}  previous work~\cite{Chatzistavrakidis:2013jqa,Kimura:2015qze,Bakhmatov:2016kfn,Blair:2017hhy}{. An important ingredient of our work will be the employment of the double field theory formalism} (DFT)~\cite{Siegel:1993th,Siegel:1993xq,Hull:2006va,Hull:2009mi,Hull:2009zb,Hohm:2010jy,Aldazabal:2013sca,Berman:2013eva,Hohm:2013bwa}.

In DFT, the T-duality symmetry O$(D,D)$ is made manifest as a space-time symmetry at the cost of doubling the number of space-time coordinates. The doubled set of coordinates are denoted by $X^M=(x^m, \tilde{x}_m)$ {with}  $x^m$ sometimes referred to as momentum coordinates and $\tilde{x}_m$ as winding coordinates. The indices $m$ take $D$ different values and $X^M$ forms a $2D$-dimensional fundamental representation of O$(D,D)$. The doubling of coordinates is a spurious operation and one must impose the O$(D,D)$ invariant \textit{section constraint}
\begin{align}
\label{eq:SC}
\eta^{MN} \partial_M \otimes \partial_N =0
\end{align}
when acting on any pair of fields on the doubled space. Here, $\eta^{MN}$ denotes the O$(D,D)$-invariant metric of split signature.

The section condition~\eqref{eq:SC} can be solved explicitly by `choosing a section', \textit{i.e.},  by making a maximal choice of coordinates among the $X^M$ on which the fields may actually depend. This could be done, for instance, by requiring that nothing depend{s} on the winding coordinates $\tilde{x}_m${. In}  this way one goes back to the usual space-time formulation. However, O$(D,D)$ acts on the coordinates $X^M$ and will therefore transform one choice of section into another.

Writing down Wess--Zumino terms in DFT requires not only to consider an embedding of the brane in doubled space-time together with appropriate space-time fields in the doubled space-time but also a choice of section. The picture we shall develop in the present paper is that, while T-duality in standard string theory often changes the dimensionality of a brane, one should think of the brane in DFT as an object of fixed dimensionality in the doubled space. The `apparent' dimensionality of a brane is then determined by the overlap of the embedded brane with the solution to the section constraint. In other words, one can use O$(D,D)$ to rotate world-volume directions out of section and thereby decrease the apparent dimensionality of the brane (or the other way around). 

For the case of D-branes, their description in a $2n$-dimensional doubled space as resulting from open strings satisfying $n$ Dirichlet and $n$ Neumann conditions, with T-duality  changing which of these directions are winding coordinates and which are momentum coordinates, and therefore changing the apparent dimensionality of the brane, was originally given in \cite{Hull:2004in} and further developed in \cite{Lawrence:2006ma,Albertsson:2008gq,Albertsson:2011ux}. In DFT, we are thus led to the interpretation of any D$(p-1)$-brane as a D9-brane where $p$ directions are momentum and the remaining $10-p$ are winding, and the aim of this paper is to derive the Wess--Zumino term for any D$(p-1)$-brane given in~\eqref{eq:WZ1} from a D9 Wess--Zumino term in DFT. 

Generalising this to all branes in string theory, we think of 
their world-volume integral  of fixed dimension for all branes of fixed type $\gs^{-\alpha}$. One can use T-duality to rotate some of the `standard` transverse directions into the winding space with the effect of creating isometry directions in the usual momentum space. This is for example the view we take on relating the NS5-brane to the Kaluza--Klein monopole, which directly follows from the analysis of the corresponding DFT background \cite{Berman:2014jsa,Bakhmatov:2016kfn}. The same picture has been shown to be true for the branes of M-theory understood as backgrounds of Exceptional Field Theory \cite{Bakhmatov:2017les,Berman:2018okd,Fernandez-Melgarejo:2018yxq}.

Using this philosophy, we can write a master Wess--Zumino term for all branes with fixed $\alpha$. The particular choice of a given brane can be implemented by fixing a `brane charge' as will be more transparent when we write down the various Wess--Zumino terms. The type of brane charge depends on the dimensionality (in doubled space) of the brane along with the DFT potential {it couples}  to. In this work we will only deal with DFT potentials that have 
a standard brane representative in ten dimensions in their U-duality orbit.
In table~\ref{tab:DFTpots}, we summarise the various DFT potentials for the different values of $\alpha$ that couple to branes. {The O$(D,D)$ representations of these potentials can be derived using  $E_{11}$~\cite{West:2001as}. They can also be extracted from~\cite{Hohm:2011zr,Bergshoeff:2016ncb,Bergshoeff:2016gub}.}
Describing gauge-invariance can require the introduction of additional O$(D,D)$ representations~\cite{Bergshoeff:2016ncb,Bergshoeff:2016gub} that partially follow from $E_{11}$ \cite{West:2010ev,Rocen:2010bk} and completely from its tensor hierarchy algebra extension~\cite{Rocen:2010bk,Greitz:2013pua,Bossard:2017wxl}.

\begin{table}[t!]
\centering
\begin{tabular}{c|c|l}
$\alpha$ & Potential & Object\\\hline
$1$ & $C_{{\alpha}}$ (spinor) & D-branes\\
$2$ & $D^{MNPQ}=D^{[MNPQ]}$ & NS-branes\\
$3$ & $E_{MN \alpha}$ ({gamma-}traceless tensor-spinor) & exotic branes containing S-dual of D7\\
$4$ & $F^+_{M_1\ldots M_{10}}=F^+_{[M_1\ldots M_{10}]}$  (self-dual) & exotic branes containing S-dual of D9\\
$4$ & $F^{M_1\ldots M_4,N_1N_2}$ ((4,2)-tensor) & exotic branes\\
$4$ & $F^{M_1\ldots M_7,N_1}$ ((7,1)-tensor) & exotic branes\\
\end{tabular}
\caption{\label{tab:DFTpots}\sl Double field theory potentials at order $\gs^{-\alpha}$ for $\alpha=1, ...,4$. BPS branes only couple to the longest weight components of these potentials~\cite{Kleinschmidt:2011vu,Bergshoeff:2013sxa}. The last two potentials do not have a standard brane representative in ten dimensions in their U-duality orbit and will not be considered in this work.}
\end{table}

Our analysis here is restricted to writing Wess--Zumino terms for $\alpha>0$. We do not discuss  the case $\alpha=0$ of the fundamental string and the Kaluza--Klein wave. In the existing literature~\cite{Hull:2006va}, the fundamental string is covered by writing down an action that is partially O$(D,D)$ invariant and in which half of the world-volume scalars are gauged away. The corresponding gauge field does not propagate in two spacetime dimensions.\footnote{For an extension of this description from DFT to Exceptional Field Theory (EFT), see  \cite{Arvanitakis:2018hfn}.} In this paper we write down the Wess-Zumino terms in a special form that are fully O(10,10) invariant and contain a charge tensor that gets rotated by the O(10,10) duality transformation. We have not been able to write any of the existing $\alpha=0$ actions available in the literature \cite{Hull:2006va} in the same form.  For this reason we begin our analysis with $\alpha>0$. For $\alpha\geq 2$ branes we shall also restrict mainly to a linearized picture for simplicity as this already brings out the most important features of our analysis.

This paper is structured as follows. In Section~\ref{sec:Dbranes} we first construct T-duality covariant and gauge invariant Wess--Zumino terms for the T-duality orbits of D-branes  for the full O$(10,10)$ DFT. We also discuss the effect of a non-zero Romans mass parameter. In Section~\ref{sec:NSbranes} we do the same for the NS5-branes.   Next, in Section~\ref{sec:exbranes} we define charges and schematically write the  covariant Wess-Zumino terms for T-duality orbits for the branes with $\gs^{-3}$ and $\gs^{-4}$. Here, we restrict ourselves to linearized O$(10,10)$ DFT. In Section~\ref{sec:concl} we make some  concluding remarks. For the convenience of the reader we have included three appendices. In Appendix~\ref{app:Odd}we summarize our notations and conventions. In Appendix~\ref{app:gtrm} we provide some details of how to derive the gauge transformation of a particular DFT potential. Finally, in Appendix~\ref{app:romans} we provide the Scherk-Schwarz ansatz for the NS-NS field $D^{MNKL}$ which incorporates Romans mass parameter in DFT. We show that the chosen ansatz leaves no dependence on the dual coordinate in the 7-form field strength.

\section{D-brane Wess--Zumino terms in DFT}
\label{sec:Dbranes}

In this section we construct Wess--Zumino (WZ) terms for D-branes in DFT. In particular, in the first subsection we consider the case of vanishing Romans mass, while in the second subsection we discuss the effect of turning on such mass parameter.
In both subsections, before studying D-brane WZ terms in DFT,
we will first review how to construct a gauge invariant WZ term for a D-brane coupled to supergravity in ten dimensions.

\subsection{Massless type IIA and type IIB}

The Ramond--Ramond (RR) potentials that are sources of D$(p-1)$-branes are $p$-forms $C_{p}$, with $p$ even in type IIB and odd in massless type IIA supergravity. We consider a democratic formulation, in which both the electric and magnetic potentials are included. In particular, in {type} IIA the potentials $C_7$ and $C_5$ are dual to $C_1$ and $C_3$, while in {type} IIB $C_8$ and $C_6$ are dual to $C_0$ and $C_2$, while $C_4$ is self-dual. On top of this, we also have a potential $C_9$ in IIA and $C_{10}$ in IIB, that are sources for D8 and D9-branes respectively. We begin with the case of zero Romans mass; massive supergravity will be treated in Section~\ref{sec:RomansD}.

We first review the standard construction of D-brane WZ terms~\cite{Aganagic:1996pe,Cederwall:1996ri,Bergshoeff:1996tu,Bergshoeff:1998ha}. Let $H_3 =d B_2$ be the field strength of the Neveu--Schwarz (NS) 2-form $B_2$. $H_3$ is gauge-invariant with respect to $\delta B_2 = d \Sigma_1$. The field strengths of the RR potentials are
\begin{equation}
G_{p+1} = d C_{p} + H_3 \wedge C_{p-2}  \qquad \quad \delta C_{p} = d \lambda_{p-1} + H_3 \wedge \lambda_{p-3}{\,,} \label{RRfieldstrengthscomponents}
\end{equation}
where we have also shown the gauge transformations{, with gauge parameters $\lambda$,} that leave {these field-strenghts}  invariant.  The RR fields defined {in} this way are invariant under {the gauge transformations with parameter} $\Sigma_1$. In order to write a gauge-invariant WZ term, one introduces a world-volume 1-form potential $b_1$, and writes its field strength as
\begin{equation}
{\cal F}_2 = d b_1 + B_2 \label{calF2}{\,,}
\end{equation}
where $B_2$ {denotes} the pull-back of the ten-dimensional NS 2-form to the world-volume of the brane.\footnote{Everywhere in the paper we will denote any supergravity potential and its pull-back with the same letter. Given that we mainly deal with brane effective actions, we assume that this will not cause any confusion. Capital Roman letters refer to space-time fields (or their pull-backs) and small letters to world-volume fields.}  In order for ${\cal F}$ to be gauge-invariant, $b_1$ has to transform under the gauge parameter $\Sigma_1$ by a shift equal to the opposite of its pull-back on the world-volume:
\begin{equation}
\delta b_1 = - \Sigma_1 \quad .
\end{equation}
The resulting gauge-invariant WZ term for a brane with charge $q$ is {given by}
\begin{equation}
q \int_{\Sigma_p} e^{{\cal F}_2 } \wedge C = q \int_{\Sigma_p} d^{p} \xi \  \epsilon^{a_1\ldots a_{p}}  \ [ e^{{\cal F}_2 } \wedge C ]_{ a_1 \ldots a_{p}}  \quad . \label{DbraneWZcomp}
\end{equation}
In this expression, the integral is {over} the world-volume coordinates $\xi^{a}$, $a=0,\ldots,p-1$, and one has to expand  $e^{{\cal F}_2 } C $ {in forms of different rank} and pick out all terms that are $p$-forms. To prove gauge invariance, one {first} integrates by parts the term that arises from the $d\lambda$ part of the variation of $C${. Next,  one can show that up to a total derivative this contribution cancels against the $H_3 \wedge \lambda$ terms. To prove this cancelation, one needs to use the fact that  eq. \eqref{calF2} implies}
\begin{equation}
\label{eq:BF}
d {\cal F}_2 = H_3 \quad ,
\end{equation}
where $H_3=dB_2$ is the pull-back on the world-volume of the NS 3-form field strength.

The aim of this section is to write down the WZ term for D-branes in a  DFT-covariant way.
In order to construct the DFT-covariant WZ term, we first review how the Ramond--Ramond potentials are described in DFT as a chiral ${\rm O}(10,10)$ spinor $C_{{\alpha}}$~\cite{Kleinschmidt:2004dy,Hohm:2011zr}. The ${\rm O}(10,10)$ Clifford algebra {is given by}
 \begin{equation}
  \{\Gamma_M,\Gamma_N\} \ = \ 2\eta_{MN}\;, \qquad
  \eta_{MN}  =  \begin{pmatrix}  0 & 1 \\ 1 & 0 \end{pmatrix} \;,
 \end{equation}
{It can be} realised in terms of fermionic oscillators as
 \begin{equation}
 \label{psiClifford}
  \{\Gamma_m,\Gamma_n\} \ = \   \{\Gamma^m,\Gamma^n\} \ = \ 0\;, \qquad \{\Gamma_m, \Gamma^n\} \ = \ 2\delta_m^{n}\;.
 \end{equation}
The split of indices here corresponds to an embedding of ${\rm GL}(10)\subset {\rm O}(10,10)${. We} should think of this as choosing the solution to the section condition in terms of the usual momentum coordinates $x^m$.  {We also observe} that
\begin{align}
\Gamma^{m_1\ldots m_p} = \Gamma^{[m_1} \cdots \Gamma^{m_p]} = \Gamma^{m_1} \cdots \Gamma^{m_p}
\end{align}
as all the gamma matrices with ${\rm GL}(10)$ upstairs indices anti-commute. More details on the O$(10,10)$ spinors are collected in Appendix~\ref{app:Odd}.

Using the relation $(\Gamma^m)^\dagger = \Gamma_m$  one observes that  the anticommutators \eqref{psiClifford} realise a fermionic harmonic oscillator.\footnote{The creation and annihilation operators are not normalised canonically but this normalisation is more convenient for writing {conventional} O$(10,10)$ spinor bilinears.}
The spinor representation is then constructed from the Clifford vacuum $|0\rangle $ satisfying
 \begin{equation}\label{CliffVacuum}
  \Gamma_m |0\rangle \ = \ 0 \quad \textrm{for all $m$.}
 \end{equation}
By taking the conjugate of this equation we also conclude that
\begin{equation}\langle 0|\Gamma^m = 0 \quad \textrm{for all $m$.}  \label{bravacuum}
\end{equation}
One then writes the RR DFT potential as\,\footnote{In this paper we always denote the DFT potentials with the same letter as the corresponding 10-dimensional potentials. From the index structure and the expressions in which these potentials occur it is always clear whether one is referring to the former or the latter{. We}  therefore  assume that this notation does not lead to confusion.}
 \begin{equation}
C   
\ =  \ \sum_{p=0}^{10}\frac{1}{p!} C_{m_1\ldots m_p}\,\Gamma^{m_1\ldots m_p} |0\rangle\;, \label{RRpotentialsinchi}
 \end{equation}
which encodes the Ramond--Ramond potentials $C_p$ of both {the} type IIA and type IIB theory, depending on whether one sums only over odd $p$ or over even $p$, corresponding to a fixed chirality of $C$. In {this}  paper we fix the chirality of $C$ to be positive, hence one recovers the right sums by imposing that in the IIA case the chirality of the Clifford vacuum is negative and in {the} IIB {case}  it is positive. {A} T-duality {transformation} corresponds to flipping the chirality of the Clifford vacuum.

We now discuss the gauge transformations of $ C$.  Defining a dressing by the NS 2-form through the Clifford element
\begin{equation}
\label{eq:SBSB}
S_B = e^{-\tfrac{1}{2} B_{mn} \Gamma^m \Gamma^n}
\quad \Rightarrow\quad
S_B \slashed{\partial} S_B^{-1} = \frac12 \partial_m B_{np} \Gamma^m \Gamma^n \Gamma^p = \frac16 H_{mnp}  \Gamma^{mnp} \,,
\end{equation}
where $\slashed{\partial}= \Gamma^M \partial_M$ with the solution to the section condition~\eqref{eq:SC} that $\tilde{\partial}^m=0$,
one can write the gauge transformation as\footnote{The DFT RR potential $C$ is related to the RR potential $\chi$ of \cite{Hohm:2011zr} by eq. \eqref{fieldredefusHK}. A more detailed analysis of the relation between the two bas{e}s in given in appendix C.}
\begin{equation}
\label{gaugeRRDFT}
\delta C  = \slashed{\partial} \lambda   + S_B \slashed{\partial} S_B^{-1} \lambda   \quad ,
\end{equation}
where the gauge parameter
\begin{align}
\lambda = \sum_{p=0}^{10} \frac{1}{p!} \lambda_{m_1\ldots m_{p}} \Gamma^{m_1\ldots m_p} |0\rangle
\end{align}
is a spinor of opposite chirality compared to $C$. The gauge-invariant DFT RR field strength is then
\begin{equation}
G  = \slashed{\partial}C  + S_B \slashed{\partial} S_B^{-1} C  = \sum_{p=0}^{10} \frac1{p!} G_{m_1\ldots m_p} \Gamma^{m_1 \ldots m_p} |0\rangle\, .
\end{equation}
It {is}  a spinor of opposite chirality compared to $C$. This field strength is also invariant under the $\Sigma_1$ gauge transformations of $B_2$ due to~\eqref{eq:SBSB}.

We want to use this notation to derive the form of the WZ term of a D-brane effective action in DFT. We will first derive the WZ term for the 9-brane in IIB, and we will then determine all the other effective actions by T-duality. The world-volume of the D9 coincides with the ten-dimensional space-time with (momentum) coordinates $x^m$. This means that the world-volume coordinates $\xi^a$  can be chosen to coincide with the coordinates $x^m$. We will only write the brane action in such adapted coordinates.\footnote{Not using this kind of static gauge would require also introducing a doubled world-volume with associated section constraint.}
In view of~\eqref{DbraneWZcomp}, we also have to include the world-volume gauge field $b_m$ in the discussion. Similarly, we define its field strength as in eq. \eqref{calF2},
in terms of which we define the gauge-invariant Clifford algebra element on the world-volume
\begin{equation}
S_{\cal F} =  e^{-\tfrac{1}{2} {\cal F}_{mn} \Gamma^m \Gamma^n} \quad . \label{defofcalF}
\end{equation}
Acting with  $S_{\cal F}$ on $C$ one obtains an expression whose gauge transformation is
\begin{equation}
\label{eq:deltachi}
\delta ( S_{\cal F}^{-1}  {C}) = S_{\cal F}^{-1} \delta C = S_{\cal F}^{-1} \slashed{\partial}    \lambda + S_{\cal F}^{-1} S_B \slashed{\partial}  S_B^{-1}  \lambda = \slashed{\partial} \left( S_{\cal F}^{-1} \lambda\right)
\end{equation}
upon using  the relation
\begin{equation}
S_{\cal F} \slashed{\partial}  S_{\cal F}^{-1} =  S_B \slashed{\partial}  S_B^{-1} \quad ,
\end{equation}
that is a consequence of eq. \eqref{eq:BF}. 
Relation~\eqref{eq:deltachi} shows that $S_{\cal F}^{-1} C$ varies into a total derivative just like~\eqref{DbraneWZcomp}. We note that by the analysis of~\cite{Hohm:2011cp}, we can extend the operator $\slashed{\partial}=  \Gamma^M \partial_M$ to range over the full doubled space which is here achieved trivially by the choice of section $\tilde{\partial}^m=0$.

Using these variables, we can rewrite the Wess--Zumino term~\eqref{DbraneWZcomp} for the case $p=9$ as
\begin{equation}
\label{D9braneWZterm}
S_{\rm WZ}^{\rm D9} = \int d^{10}\xi \  \overline{Q}_{10}  S_{\cal F}^{-1}  C
\end{equation}
where
\begin{align}
\label{eq:Q10}
\overline{Q}_{10} &= \frac{q}{2^{10}} \langle 0 | \Gamma_0 \cdots \Gamma_9  \quad .
\end{align}
As already mentioned, in this expression the world-volume coordinates coincide with the coordinates $x^m$. We want to show that the other D-branes arise from the action of T-duality on this expression. The effect of T-duality is to rotate the charge of the brane so that its world-volume starts invading the $\tilde{x}$ space. This is what we are going to discuss in the following.

\begin{figure}
\centering
\begin{tikzpicture}
\draw[line width=1pt] (0,2.75) -- (10,2.75) -- (10,3.25) -- (0,3.25)--cycle;
\draw[line width=1pt] (5,2.75) -- (5,3.25);
\node[align=center] at (2.5,3) {momentum $x^m$};
\node[align=center] at (7.5,3) {winding $\tilde{x}_m$};
\node[align=center] at (-1.9,3) {Doubled space-time};
\draw[pattern=north west lines,line width=1pt] (0,2) -- (5,2) -- (5,2.5) -- (0,2.5)--cycle;
\node[align=center] at (-0.5,2.25) {D9};
\node (1) at (-0.7,2.25) {};
\draw[line width=1pt] (0.5,1.25) -- (5.5,1.25) -- (5.5,1.75) -- (0.5,1.75)--cycle;
\draw[pattern=north west lines,line width=0.1pt] (0.5,1.25) -- (5,1.25) -- (5,1.75) -- (0.5,1.75)--cycle;
\node[align=center] at (-0.5,1.5) {D8};
\node (2) at (-0.7,1.5) {};
\draw[->,line width=1pt] (1) to [out=-180,in=180,looseness=2] (2);
\node[align=center] at (-2.25,1.875) {T-duality};
\end{tikzpicture}
\caption{\label{fig:Tdual}\textit{D-branes in doubled space. All branes have a ten-dimensional world-volume and the intersection of this with the ten physical momentum dimensions gives the apparent dimensionality of the world-volume. T-duality along an isometry direction can move part of the ten-dimensional world-volume between momentum and winding directions.}}
\end{figure}
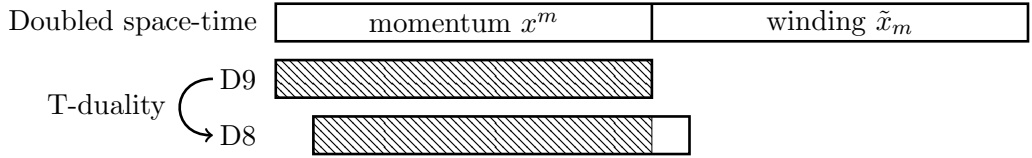

To understand to what extent~\eqref{D9braneWZterm} can also be used for the other D$(p-1)${-branes} we consider the effect of a T-duality transformation along a world-volume direction of the D9-brane, leading to a D8-brane. We will be describing the T-duality in a way where we still think of the momentum directions $x^m$ as the physical ones and keeping the form of $C$ as in~\eqref{RRpotentialsinchi} but rather transform the brane by acting on its charge. If the T-duality transformation is performed along the $9$-direction, say, then the brane no longer extends along the momentum direction $x^9$ but rather along the winding direction $\tilde{x}_9$. This is shown in Figure~\ref{fig:Tdual}. For T-duality $\tilde{x}_9$ is an isometry direction, which also follows from the strong constraint. Let us denote the charge obtained after T-duality by $\overline{Q}_9$. It equals
\begin{align}
\label{eq:Q9}
\overline{Q}_9 = \frac{q}{2^9} \langle 0| \Gamma_0 \cdots \Gamma_8 = \overline{Q}_{10} \Gamma^9\,,
\end{align}
which shows how T-duality acts on {the} charges. This transformed charge has the property that
\begin{align}
\overline{Q}_9 S_{\cal F}^{-1} C = [e^{{\cal F}_2} \wedge C]_{\textrm{$9$-form on world-volume}}
\end{align}
and so projects to the {correct} RR potential {that is appropriate for describing} the WZ term of a D8-brane. While the charge~\eqref{eq:Q10} is invariant under the ${\rm SO}(1,9)$ of the momentum directions, the charge $Q_9$ is only invariant under its subgroup ${\rm SO}(1,8)$.

The WZ term obtained by T-duality of~\eqref{D9braneWZterm} is thus {given by}
\begin{align}
S_{\rm WZ}^{\rm D8} = \int d^{10} \xi \overline{Q}_9 S_{\cal F}^{-1} C = \int d^{9} \xi \overline{Q}_9 S_{\cal F}^{-1} C\,.
\end{align}
This integral is initially over ten dimensions. But, as argued above, the direction $\tilde{x}_9$ that is now part of the ten world-volume directions is an isometry and hence nothing in the integral depends on it. We can thus perform this integral and, for a proper normalisation, simply obtain the correct nine-dimensional world-volume integral for the D8-brane.

The overall picture following from these considerations is that the general D-brane Wess--Zumino term is given by
\begin{align}
\label{eq:DpDFT}
S_{\rm WZ}^{{\rm D}(p-1)} =  \int d^{10}\xi \   \ \overline{Q}_{p}  S_{\cal F}^{-1} C
\end{align}
and thus always {involves an integral that is formally} ten-dimensional. {It}  is understood {here} that $\overline{Q}_p$ consists of the O$(10,10)$ gamma matrices that characterise the intersection of the ten-dimensional world-volume with the ten physical momentum direction{s}. We must think of any D-brane as a 9-brane, where some of its world-volume directions {have invaded} the winding space. The information of how many directions are momentum and how many are winding is carried by the charge $Q_p$, and T-duality acts on this charge.

\subsection{Massive type IIA supergravity}
\label{sec:RomansD}

We now return to the issue of allowing the Romans mass to be different from zero in {type} IIA {supergravity}. The Romans mass modifies the field strengths \eqref{RRfieldstrengthscomponents} and their gauge transformations as follows~\cite{Romans:1985tz}:
\begin{equation}
G_{p+1} = d C_{p} + H_3 \wedge C_{p-2} + m e^{-B_2}{\,,} \qquad \quad \delta C_{p} = d \lambda_{p-1} + H_3 \wedge \lambda_{p-3} + m \Sigma_1 \wedge
e^{-B_2}\,,
\label{RRfieldstrengthscomponentsRomans}
\end{equation}
where we recall that $\Sigma_1$ is the gauge parameter of $B_2$.
As a consequence, the gauge-invariant WZ term takes the modified form
\begin{equation}
\label{DbraneWZcompRomans}
\int [ e^{{\cal F}_2 } \wedge C + m b_1 \wedge \tfrac{1}{f_2}( e^{f_2}-1) ]  \, ,
\end{equation}
where gauge-invariance requires the inclusion of the additional Chern--Simons term~\cite{Bergshoeff:1996cy,Green:1996bh}, and $f_2=db_1$ {is the field-strength ${\cal F}_2$ of $b_1$ without the inclusion of $B_2$}. The Chern--Simons term has the property that
\begin{align}
d \left(b_1 \wedge \frac{1}{f_2} ( e^{f_2}-1) \right)  = e^{f_2}-1 = \sum_{k\geq 1} \frac{1}{k!} f_2^{k}\,.
\end{align}

We want to {recast the above expressions within}  DFT. The closure of the gauge transformations~\eqref{gaugeRRDFT} actually allows for a mild violation of the strong constraint~\cite{Hohm:2011cp}. The procedure is  similar to a generalised Scherk--Schwarz mechanism, in which the RR DFT potential becomes~\cite{Hohm:2011cp}
\begin{equation}
C
\quad\longrightarrow \quad
 C + \frac{m}{2} S_B \tilde{x}_1 \Gamma^1  | 0\rangle \, . \label{hohmkwakansatz}
\end{equation}
Here, we have introduced a mild linear $\tilde{x}_1$ dependence; the choice of $\tilde{x}_1$ is completely arbitrary and nothing depends on choosing this particular direction.
The field strength associated with this
is then
\begin{equation}
\label{eq:Gmass}
G = \slashed{\partial} C  + S_B \slashed{\partial} S_B^{-1} C  + m S_B | 0\rangle \, .
\end{equation}
This field strength is gauge-invariant if the RR potentials $C$ also transform with a St\"uckel\-berg shift under the $B_2$ gauge parameter $\Sigma_1$ as~\cite{Hohm:2011cp}
\begin{equation}
\label{eq:gaugemass}
\delta_\Sigma C  = m S_B \Sigma_m \Gamma^m | 0 \rangle \,.
\end{equation}
This can be seen by
\begin{align}
\delta_\Sigma G =  m \slashed{\partial} (S_B \Sigma_m \Gamma^m) |0\rangle + m S_B \slashed{\partial} S_B^{-1}  S_B \Sigma_m \Gamma^m |0\rangle - m S_B \partial_m \Sigma_n \Gamma^m \Gamma^n |0\rangle =0
\end{align}
upon using the identity
\begin{equation}
S_B \slashed{\partial} S_B^{-1}  S_B = - \slashed{\partial} S_B \label{SBslashSBslash}
\end{equation}
together with  $\slashed{\partial}\Sigma_m \Gamma^m = \partial_m \Sigma_n \Gamma^m \Gamma^n$.
Equations~\eqref{eq:Gmass} and~\eqref{eq:gaugemass} reproduce {the transformations} \eqref{RRfieldstrengthscomponentsRomans}.

The WZ term~\eqref{DbraneWZcompRomans} can then be written in DFT by replacing $S_{\cal F}^{-1} C$ in~\eqref{eq:DpDFT} by
\begin{equation}
  S_{\cal F}^{-1}C \quad \longrightarrow \quad
  S_{\cal F}^{-1}C \ +
m \frac{1}{2^n} \frac{1}{(n+1)!} b_{a_1} f_{a_2 a_3} \cdots f_{a_{p-1} a_p} \Gamma^{a_1} \cdots \Gamma^{a_p} | 0 \rangle\,,
\end{equation}
where $n = \frac{p-1}{2}$.

\section{NS-brane WZ terms in DFT}
\label{sec:NSbranes}

In this section we discuss how the NS5-brane WZ term {and its T-dual partner branes are} written in DFT. As in the previous section, we fi{r}st discuss the massless IIA and IIB theories, and we then discuss the WZ term in {the} massive IIA theory.

\subsection{Massless type IIA and type IIB}
We first write down the WZ term of the NS5{-brane} in supergravity, for both the {massless} IIA and IIB theory, as done in \cite{Bergshoeff:2011zk}.
The NS5-brane is electrically charged under {the 6-form potential} $D_6$, which is the  magnetic dual of the NS 2-form potential $B_2$. The gauge transformation of $D_6$ in the {massless} IIA theory is {given by}
\begin{equation}
\delta D_6 = d \Xi_5 + \lambda_0 \ G_6 - \lambda_2 \wedge G_4 + \lambda_4 \wedge G_2  \label{D6gaugetransfIIA}
\end{equation}
and the corresponding gauge invariant field strength {reads}
\begin{equation}
H_7 =d D_6 - C_1 \wedge G_6 + C_3 \wedge G_4 - C_5 \wedge G_2 \quad .\label{H7fieldstrengthIIA}
\end{equation}
In the IIB theory, the gauge transformation is {given by}
\begin{equation}
\delta D_6  = d \Xi_5 + \lambda_1 \wedge G_5 - \lambda_3 \wedge G_3 + \lambda_5\wedge  G_1 \label{D6gaugetransfIIB}
\end{equation}
and the field strength {reads}\,\footnote{The term $C_0 \ G_7$ is not required by gauge invariance but is a consequence of T-duality, as will become more clear from the DFT analysis. The same applies to the term ${\cal G}_6 \ C_0$ in eq. \eqref{WZNS5IIB}.}
\begin{equation}
H_7 = d D_6 + C_0 \ G_7 - C_2 \wedge G_5 + C_4 \wedge G_3 - C_6 \wedge G_1 \quad . \label{H7fieldstrengthIIB}
\end{equation}
To show {the} gauge invariance of $H_7$ in both theories one has to use the Bianchi identities
\begin{equation}
d G_{p+1} = - H_3 \wedge G_{p-1}
\end{equation}
which follows from {eq.~}\eqref{RRfieldstrengthscomponents}.
To construct the WZ term, one introduces the world-volume potentials $c_{p-1}$, whose gauge invariant field strengths are
\begin{equation}
{\cal G}_p = d c_{p-1} + C_p + H_3 \wedge c_{p-3}
\end{equation}
satisfying the Bianchi identity
\begin{equation}
d {\cal G}_p = G_{p+1} - H_3 \wedge {\cal G}_{p-2} \quad .\label{BianchicalG}
\end{equation}
Again, as in the previous section, in these expressions it is understood that all the supergravity fields are {pulled-back to} the six-dimensional world-volume of the NS5-brane. In addition, all these field-strengths satisfy the duality relations
\begin{equation}
\label{GhodgeG}
{\cal G}_p = *_6 {\cal G}_{6-p}
\end{equation}
on the world-volume, which in particular implies that in the IIA case the world-volume potential $c_2$ is self-dual \cite{Bergshoeff:2011zk}.
In order for {the world-volume field-strengths} ${\cal G}$ to be gauge invariant, the world-volume potentials have to shift by the opposite of the pull-back on the world-volume of the RR gauge parameters,
\begin{equation}
\delta c_{p-1} = - \lambda_{p-1}\quad .
\end{equation}

We now want to use these ingredients to construct {gauge-invariant} WZ terms.
One finds that the WZ term  for the NS5-brane in the IIA theory is {given by}
\begin{equation}
\int [ D_6 - {\cal G}_1 \wedge C_5 + {\cal G}_3 \wedge C_3 - {\cal G}_5 \wedge C_1 ] \quad .\label{WZNS5IIA}
\end{equation}
{Similarly,  the WZ term in the IIB theory reads}
\begin{equation}
\int [ D_6 - {\cal G}_0 \ C_6 + {\cal G}_2 \wedge C_4 - {\cal G}_4\wedge  C_2 + {\cal G}_6 \ C_0] \quad . \label{WZNS5IIB}
\end{equation}
In this latter expression one needs the auxiliary 0-form ${\cal G}_0$, that is not the field-strength of any world-volume potential but satifies the Bianchi identity
\begin{equation}
d {\cal G}_0 = G_1
\end{equation}
which is a particular case of eq. \eqref{BianchicalG} and whose solution is simply ${\cal G}_0 = C_0$.
In {the IIB} case the world-volume fields are a vector $c_1$ and its dual $c_3$.

We {wish}  to discuss what happens to this brane under T-duality. If the T-duality is along the world-volume, this maps the NS5{-brane} of one theory to the NS5{-brane} of the other theory. On the other hand, if the T-duality is along a transverse direction,  {the NS5-brane of one theory is mapped to} the KK monopole of the other theory. This generalises if one keeps performing T-dualities in transverse directions. In particular, a further T-duality leads to the $5_2^2$ brane of \cite{deBoer:2010ud}, and proceeding this way one obtains the non-geometric branes $5_2^3$ and $5_2^4$. Denoting with $5_2^0$ and $5_2^1$ the NS5{-brane}  and KK monopole, this is summarised by the chain
\begin{equation}
5_2^0 \leftrightarrow 5_2^1 \leftrightarrow 5_2^2 \leftrightarrow 5_2^3  \leftrightarrow 5_2^4 \quad .
\end{equation}
We want to {reproduce this behaviour under T-duality}  from a DFT formulation of the WZ term of the NS5-brane. In order to achieve this, we first discuss how the $D_6$ potential {can be seen as a particular component of} a DFT potential.

The ${\rm O}(10,10)$ potential that contains $D_6$ is the field $D^{MNPQ}$ in the completely antisymmetric representation with four indices.
In particular, $D_6$ is the potential that results from contracting the component with all upstairs indices $D^{mnpq}$ with the ten-dimensional epsilon symbol. The other components $D^{mnp}{}_q$, $D^{mn}{}_{pq}$, $D^{m}{}_{npq}$ and $D_{mnpq}$ correspond instead to
the mixed-symmetry potentials $D_{7,1}$, $D_{8,2}$, $D_{9,3}$ and $D_{10,4}$, associated to the  $5_2^1$, $5_2^2 $, $5_2^3$ and $5_2^4$ brane respectively, together with the potentials $D_8$, $D_{9,1}$, $D_{10,2}$ and $D_{10}$.\footnote{Here and in the rest of the paper all mixed-symmetry potentials belong to irreducible representations of ${\rm SL}(10,\mathbb{R})$. This means that for instance $D_{7,1}$ is the traceless part of $D^{mnp}{}_q$, while $D_8$ is its trace, and similarly for the other cases. The potentials coming from traces of the DFT potential do not couple to branes.}
  As shown in \cite{Bergshoeff:2016ncb}, the action for the potential $D^{MNPQ}$ arises from dualising the linearised DFT action, but this can only be achieved if one also introduces the additional auxiliary potentials $D^{MN}$ (with indices antisymmetrised) and $D$. The fields $D^{MNPQ}$ and $D$ appear in the $E_{11}$ while $D^{MN}$ is only part of its tensor hierarchy extension. The field equations contain all these potentials via the field strengths
\begin{align}
& H^{MNP} = \partial_Q D^{QMNP} + 3 \partial^{[M} D^{NP ]}\,, \nonumber \\
& H^M = \partial_N D^{NM} + \partial^M D \quad ,\label{fieldstrengthsofDsDFT}
\end{align}
that are invariant under the gauge transformations
\begin{align}
& \delta D^{MNPQ} =\partial_R \Xi^{RMNPQ} + 4 \partial^{[M} \Xi^{NPQ]}\,, \nonumber \\
& \delta D^{MN} =\partial_P \Xi^{PMN} + 2 \partial^{[M} \Xi^{N]}\,, \\
& \delta D = \partial_M \Xi^M \quad . \nonumber
\end{align}

We now want to see whether one can add to the field strengths in eq. \eqref{fieldstrengthsofDsDFT} non-linear couplings to the RR potentials. More precisely, we want to add to $H^{MNP}$ and $H^M$ the terms $\bar{G} \Gamma^{MNP} C$ and $\bar{G} \Gamma^M C$, and to the gauge transformations of $D^{MNPQ}$, $D^{MN}$ and $D$ the terms $\bar{G} \Gamma^{MNPQ} \lambda$, $\bar{G} \Gamma^{MN} \lambda$ and $\bar{G} \lambda$.\footnote{The conventions for the ${\rm O}(10,10)$ spinor bilinears are discussed in Appendix \ref{app:Odd}.} It turns out that this is impossible: there is no set of coefficients {for} these terms that gives a gauge invariant field strength. The terms that cannot be cancelled are the ones  in which either $G$ or $\lambda$ are hit by a derivative carrying a non-contracted index. This means that, because of the section condition, such terms vanish as long as the corresponding index is upstairs. The outcome of this analysis is that one can only write down gauge invariant couplings to the RR potentials for the field strength with all upstairs indices $H^{mnp}$, which gives the field strength $H_7$ of $D_6$ by contraction with an epsilon symbol. This is  consistent with the fact that only for $D_6$ the dualisation procedure works at the full non-linear level.

Keeping in mind the analysis above, we can write down the gauge transformations  of $D^{MNPQ}$ as
\begin{equation}
\delta D^{MNPQ} = \partial_R \Xi^{RMNPQ} + \overline{G} \Gamma^{MNPQ} \lambda \label{DMNPQgaugetransffull}	
\end{equation}
and study how the field strength
\begin{equation}
H^{MNP} = \partial_Q D^{QMNP} +  \overline{G} \Gamma^{MNP} C \label{HMNPDFTfieldstrength}
\end{equation}
transforms.\footnote{Given the analysis above, there is no need to consider the parameter $\Xi^{MNP}$, the potential  $D^{MN}$ and its variation in $H^{MNP}$ because for the component $H^{mnp}$ these terms vanish due to the section condition.}
 Using the Bianchi identity for $G$,
\begin{equation}
{\partial\mkern-10mu/} G =-S_B {\partial\mkern-10mu/}  S_B^{-1} \ G \quad ,
\end{equation}
which can be derived using eq. \eqref{SBslashSBslash},
one can prove,  as anticipated,  that the variation of $H^{MNP}$ vanishes up to terms  in which the index of the derivative is a free index.
In Appendix~\ref{app:gtrm} we show that for the components $D^{mnpq}$ and $H^{mnp}$  eqs. \eqref{DMNPQgaugetransffull} and \eqref{HMNPDFTfieldstrength} reproduce eqs. \eqref{D6gaugetransfIIA} and \eqref{H7fieldstrengthIIA} in the IIA case and \eqref{D6gaugetransfIIB} and \eqref{H7fieldstrengthIIB} in the IIB case.

We now want to use these results to write WZ terms. To do this, we want to get the DFT equivalent of the analysis performed at the beginning of this section.  First of all,  we observe that once one identifies the six world-volume directions with six of the $x$'s, there remains an ${\rm O} (4,4)$ subgroup of ${\rm O}(10,10)$ that rotates the transverse directions in DFT. More precisely, the brane breaks ${\rm O}(10,10)$ to ${\rm O}(6,6) \times {\rm O}(4,4)$. The ${\rm O}(10,10)$ gamma matrices decompose as
\begin{equation}
\Gamma_M  = ( \Gamma_A , \Gamma_{\hat{M}} \Gamma^* )\,,
\end{equation}
where $\Gamma_A$ are the ${\rm O}(6,6)$ gamma matrices, $\Gamma^*$ is the ${\rm O}(6,6)$ chirality matrix and $\Gamma_{\hat{M}}$ are the ${\rm O}(4,4)$ gamma matrices. The RR spinor $C$ belongs to the  spinor representation ${\bf 512}_S$ which decomposes as
\begin{equation}
{\bf 512}_S = ({\bf 8}_S , {\bf 32}_S ) \oplus ({\bf 8}_C , {\bf 32}_C ) \quad .
\end{equation}
The conjugate ${\bf 512}_C$ representation decomposes instead as
\begin{equation}
{\bf 512}_C = ({\bf 8}_S , {\bf 32}_C ) \oplus ({\bf 8}_C , {\bf 32}_S ) \quad .
\end{equation}
The world-volume potentials describing the D-branes ending on the NS5-brane collect in the spinor $c_{\dot{\alpha}}$ in the ${\bf 512}_C$, transforming as
\begin{equation}
\delta c = - \lambda\,.
\end{equation}
{One can define a} gauge-invariant world-volume field strength {${\cal G}$ as}
\begin{equation}
{\cal G} = {\partial\mkern-10mu/} c + C + S_B {\partial\mkern-10mu/} S_B^{-1} c\,,
\end{equation}
where $\ddt = \Gamma^A \partial_A$.
The field strength ${\cal G} $ satisfies the Bianchi identity
\begin{equation}
{\partial\mkern-10mu/} {\cal G} = G - S_B {\partial\mkern-10mu/} S_B^{-1} {\cal G} \quad .\label{dslashcalG}
\end{equation}
One can {now} try to write down the DFT fields that occur in the Wess--Zumino term using the transverse gamma matrices as
\begin{equation}
D^{\hat{M}\hat{N}\hat{P}\hat{Q}} +  \overline{\cal G} \Gamma^{\hat{M}\hat{N}\hat{P}\hat{Q}} C \quad , \label{DMNPQWZ}
\end{equation}
whose gauge transformation is
\begin{equation}
\delta\left(D^{\hat{M}\hat{N}\hat{P}\hat{Q}} +  \overline{\cal G} \Gamma^{\hat{M}\hat{N}\hat{P}\hat{Q}} C \right) = \overline{G} \Gamma^{\hat{M}\hat{N}\hat{P}\hat{Q}} \lambda + \overline{\cal G} \Gamma^{\hat{M}\hat{N}\hat{P}\hat{Q}} {\partial\mkern-10mu/} \lambda + \overline{\cal G} \Gamma^{\hat{M}\hat{N}\hat{P}\hat{Q}} S_B {\partial\mkern-10mu/} S_B^{-1} \lambda  \quad .
\end{equation}
Integrating by part the second term we get up to a total derivative
\begin{equation}
\overline{G} \Gamma^{\hat{M}\hat{N}\hat{P}\hat{Q}} \lambda  - \partial_{\hat{R}} \overline{\cal G} \Gamma^{\hat{R}} \Gamma^{\hat{M}\hat{N}\hat{P}\hat{Q}}  \lambda - 8 \partial^{\hat{Q}} \overline{\cal G} \Gamma^{\hat{M}\hat{N}\hat{P}}  \lambda
 + \overline{\cal G} \Gamma^{\hat{M}\hat{N}\hat{P}\hat{Q}} S_B {\partial\mkern-10mu/} S_B^{-1} \lambda  \quad .
\end{equation}
Using eq. \eqref{dslashcalG} one can show that the second term cancels  with  the first  and the last {term} up to terms containing a  derivative  with respect to a free index. Similarly, the third term, which also contains a derivative with respect to a free index, does not cancel. We therefore must impose that these terms vanish. Decomposing the index of the derivative in upstairs and downstairs indices of ${\rm GL}(10,\mathbb{R})$,  this happens either because of the section condition if the free index is upstairs, or because the free index corresponds to an isometry direction if the index is downstairs. In the case of the NS5-brane we clearly are in the former situation, because as we already mentioned this corresponds to the component $D^{\hat{m} \hat{n} \hat{p} \hat{q}}$.

We {now} write the WZ term of the NS5{-brane}  as
\begin{equation}
S_{WZ}^{\rm NS5} = \int d^{6}\xi \  Q_{\hat{M}\hat{ N}\hat{P}\hat{Q} } [  D^{\hat{M}\hat{N}\hat{P}\hat{Q}} +  \overline{\cal G} \Gamma^{\hat{M}\hat{N}\hat{P}\hat{Q}} C ]  \quad . \label{NS5braneWZtermgaugefixed}
\end{equation}
Although the expression appears {to be} covariant under ${\rm O}(4,4)$, we should remember that it is only gauge invariant for the charge component
$Q_{\hat{m}\hat{n}\hat{p}\hat{q}}$ with all indices down, corresponding to the NS5-brane. In particular, expanding the bilinear in a way analogous to the analysis in Appendix~\ref{app:gtrm}  one can show that this expression gives either eq. \eqref{WZNS5IIA} or \eqref{WZNS5IIB}, according to the choice of the chirality of the Clifford vacuum.
If one performs a T-duality along a world-volume direction, this apparently does not do anything to eq. \eqref{NS5braneWZtermgaugefixed}, but this is actually not true, because T-duality is flipping the chirality of the Clifford vacuum, so that the NS5{-brane} of one theory is mapped to the NS5{-brane} of the other theory. In what follows we will discuss what happens
if one instead performs a T-duality transformation along the transverse directions.

Starting from  the charge $Q_{\hat{m}\hat{n}\hat{p}\hat{q}}$ and T-dualising along $\hat{q}$, one ends up with the charge $Q_{\hat{m}\hat{n}\hat{p}}{}^{\hat{q}}$. This corresponds to the WZ term for the potential $D^{\hat{m}\hat{n}\hat{p}}{}_{\hat{q}}$, but using eq. \eqref{DMNPQgaugetransffull} one can show that the WZ term \eqref{NS5braneWZtermgaugefixed} is no longer gauge invariant, and more precisely the non-vanishing terms in its gauge variation contain derivatives with respect to $\hat{q}$, which is no longer zero using the section condition because  $\hat{q}$ is now a downstairs index, \textit{i.e.}, the derivative is with respect to a coordinate $x$.
This means that one has to assume {that} the $x^{\hat{q}}$ is an isometry direction, and if one does that, then eq. \eqref{NS5braneWZtermgaugefixed} with this charge gives the gauge invariant WZ term for the KK monopole.\footnote{We also require for consistency that the  gauge parameters do not depend on $x^{\hat{q}}$ , which implies in particular that the gauge parameter $\Xi^{\hat{m}\hat{n}\hat{p}}$ does not contribute to the variation of $D^{\hat{m}\hat{n}\hat{p}}{}_{\hat{q}}$.}
It is important to observe that from the point of view of our analysis there is no difference between the KK monopole with one isometry along $x^{\hat{q}}$ and the NS5{-brane} with transverse coordinate $\tilde{x}_{\hat{q}}${. The} condition that the $x^{\hat{q}}$ is isometric, which means that the fields do not depend on such coordinate, is equivalent to the condition that for the rotated NS5 the coordinate $x^{\hat{q}}$ is no longer a transverse coordinate because is is replaced by  $\tilde{x}_{\hat{q}}$, on which nothing depends because of the section condition.
This can be generalised if more than one index of the charge is upstairs. All upstairs indices correspond to isometry directions for the brane.

 The final outcome of this analysis is that in general we can interpret any $\alpha=2$ brane as an NS5{-brane where some of the}   transverse directions {have invaded the}  $\tilde{x}$ space, and the expression \eqref{NS5braneWZtermgaugefixed} gives the Wess-Zumino for all such branes and their coupling to the D-branes.

\subsection{Massive type IIA supergravity}

We now finally come to the issue of the Romans mass $G_0 =m$. We first discuss how the gauge transformation of $D_6$ gets modified and how this induces additional couplings in the WZ term. Then we will move on to discussing how this is realised in DFT. When $m\neq 0$, the gauge transformation of the 6-form potential $D_6$ becomes
\begin{equation}
\delta D_6 = d \Sigma_5 + G_6 \lambda_0 - G_4 \wedge \lambda_4 + G_2 \wedge \lambda_4 - G_0 \  \lambda_6 - m \lambda \wedge e^{B_2} -m\Sigma_1  \wedge C \wedge  e^{B_2}\, ,\label{deltaD6Romans}
\end{equation}
where the field strengths $G$ are defined in eq. \eqref{RRfieldstrengthscomponentsRomans}. The gauge invariant field strength is {given by}
\begin{equation}
H_7 = d D_6 - G_6 \wedge C_1 + G_4  \wedge C_3 - G_2 \wedge C_5 + G_0 \  C_7 + m C\wedge e^{B_2}\quad .\label{H7Romans}
\end{equation}
{Furthermore, one finds that} the world volume gauge invariant field strengths are {given by} (here $b_1$ is the world-volume vector, the one that occurs in ${\cal F}_2= d b_1 + B_2$)
\begin{align}
& {\cal G}_1= d c_0 + C_1 +  m b_1\,,\nonumber \\
& {\cal G}_3 = d c_2 + H_3 \ c_0 + C_3 - \frac{1}{2} m b_1 \wedge  B_2 -\frac{1}{2} m b_1 \wedge {\cal F}_2\,, \label{calGRomans}\\
& {\cal G}_5 = d c_4 + H_3 \wedge c_2 + C_5 + \frac{1}{6} m b_1 \wedge B_2^2 + \frac{1}{6} m b_1 \wedge B_2\wedge  {\cal F}_2 + \frac{1}{6} m b_1 \wedge {\cal F}_2^2\,.\nonumber
\end{align}
{The}  gauge transformations of the world-volume fields with respect to the bulk gauge parameters are {given by}
 \begin{align}
 & \delta c_0 = -\lambda_0\,, \nonumber \\
 & \delta c_2 = -\lambda_2 +\frac{1}{2} m \Sigma_1\wedge  b_1\,, \label{gaugetransfofwvcromans}
\\
& \delta c_4 = -\lambda_4 -\frac{1}{3} m \Sigma_1 \wedge b_1 \wedge B_2 - \frac{1}{6} m \Sigma_1\wedge  b_1\wedge  {\cal F}_2\,. \nonumber
 \end{align}
One can check that {with these rules} the field strengths in eq. \eqref{calGRomans} are gauge invariant.
We find that,  in order to obtain a fully gauge invariant WZ term, one has to consider also the world volume 6-form potential $c_6$ whose gauge transformation reads
\begin{equation}
\delta c_6 = -\lambda_6 + \frac{1}{8}m \Sigma_1 \wedge b_1 \wedge B_2^2 + \frac{1}{12}m \Sigma_1 \wedge b_1 \wedge B_2 \wedge {\cal F}_2+ \frac{1}{24 }m \Sigma_1\wedge  b_1\wedge  {\cal F}_2^2 \quad .\end{equation}
Formally{,}  one can show that this transformation is exactly the one that would make the field strength
\begin{equation}
{\cal G}_7 = d c_6 + H_3 \wedge c_4 + C_7 - \frac{1}{24} m b_1 \wedge [ B_2^3 +  B_2^2 \wedge {\cal F}_2 +  B_2 \wedge {\cal F}_2^2 +{\cal F}_2^3 ]
\end{equation}
gauge invariant.\footnote{This is only formal because {in 6 dimensions} this field strength vanishes identically.}

{Putting everything together, we} find that {a} gauge invariant WZ term is {given by}
\begin{equation}
\int [ D_6 - {\cal G}_1 \wedge C_5 + {\cal G}_3 \wedge C_3 - {\cal G}_5 \wedge C_1 - m c \wedge e^{B_2 } - m c \wedge e^{{\cal F}_2}]\quad .
\end{equation}
To show gauge invariance, one has to use the Bianchi identities
\begin{equation}
d {\cal G} = G - H_3 \wedge {\cal G} - m e^{{-\cal F}_2} \label{BianchicalGRomans}
\end{equation}
which can be proven by direct computation from eqs. \eqref{calGRomans}.

We now want to recover these results in DFT. In Appendix~\ref{app:romans} we show that the field strength $H^{MNP}$ in the presence of the Romans mass arises from a generalised Scherk-Schwarz ansatz as in eq. \eqref{SSansatzDMNPQDMN}. The final result is eq.~\eqref{finalHromansappendix}, which can be written, after  performing the field redefinition of \eqref{fieldredefusHK} as
\begin{equation}
H^{MNP} = \partial_Q D^{QMNP} + \overline{C} \Gamma^{MNP} G + m \overline{C} \Gamma^{MNP} S_B | 0\rangle \quad ,
\end{equation}
which reproduces eq.~\eqref{H7Romans}. The gauge transformation of the potential $D^{MNPQ}$ is
\begin{align}
\delta D^{MNPQ} & = \partial_R \Xi^{RMNPQ} +\overline{G} \Gamma^{MNPQ} \lambda \nonumber \\
& - m \overline{\lambda} \Gamma^{MNPQ} S_B | 0\rangle - m \Sigma_R \overline{C} \Gamma^{RMNPQ} S_B | 0\rangle  \quad ,
\end{align}
{reproducing} eq.~\eqref{deltaD6Romans}.

To write down the NS5 WZ term for $m\neq 0$, we need the DFT expression for the world-volume field strengths in eq.~\eqref{calGRomans}. One finds
\begin{align}
{\cal G} &= {\partial\mkern-10mu/} c + C + S_B {\partial\mkern-10mu/} S_B^{-1} c \nonumber\\
&+ m b_A \Gamma^A \sum_{N=0}^3 \frac{1}{N+1} \left( S_B^{(N)} + \frac{1}{N} \sum_{n=1}^{N-1} S_B^{(N-n)} S_{\cal F}^{(n)} + S_{\cal F}^{(N)}  \right) | 0\rangle \quad ,
\end{align}
where with $S_B^{(n)}$ we mean the term at order $n$ in the expansion {of $S_B$} in terms of $B$ (and similarly for $S_{\cal F}$). The world-volume field strength ${\cal G}$ satisfies the Bianchi identity
\begin{equation}
{\partial\mkern-10mu/} {\cal G} =  G - S_B {\partial\mkern-10mu/} S_B^{-1} {\cal G} - m S_{\cal F}| 0\rangle \quad .
\end{equation}
Using these results, one finally finds the {following expression for the} WZ term{:}
\begin{equation}
S_{WZ}^{{\rm NS5}m} = \int d^{6}\xi \  Q_{\hat{M}\hat{ N}\hat{P}\hat{Q} } [  D^{\hat{M}\hat{N}\hat{P}\hat{Q}} +  \overline{\cal G} \Gamma^{\hat{M}\hat{N}\hat{P}\hat{Q}} C  -m \overline{c} \Gamma^{\hat{M}\hat{N}\hat{P}\hat{Q}} (S_B + S_{\cal F} )  | 0\rangle ] \quad . \label{NS5braneWZtermgaugefixedRomans}
\end{equation}
Starting from this action with charge $Q_{\hat{m}\hat{n}\hat{p}\hat{q}}$, corresponding to the NS5-brane in the presence of a Romans mass parameter, one can obtain {the} other WZ terms  in the T-duality orbit precisely as discussed in the massless case.

\section{WZ term for other exotic branes in DFT}
\label{sec:exbranes}

In the previous two sections we have shown how the WZ terms of D-branes and NS-branes can be written in a DFT-covariant way. The WZ term is contracted with a charge, and T-duality corresponds to a rotation of the charge in DFT. We have seen how for the case of D-branes a rotation of the charge gives a rotation of the embedding coordinates in double space. As a result, we can think of any D$(p-1)$-brane as a D9-brane in which $10-p$ world-volume coordinates invade the tilde space and thus become isometry directions. In the case of NS-branes, T-duality along the transverse directions also rotates them in tilde space, and thus for instance a KK monopole can be thought of as {a} NS5-brane with one direction along $\tilde{x}$.

In this section we discuss additional branes, that are the S-dual of the D7-brane and the S-dual of the D9-brane in {the type} IIB {theory}.   In the first subsection we discuss the S-dual of the D7-brane. This brane has a tension scaling like $g_S^{-3}$, and it is related by T-duality to a chain of exotic branes as discussed in \cite{Bergshoeff:2011ee}. In the second subsection we discuss the branes related by T-duality to the S-dual of the D9{-brane}.

\subsection{\texorpdfstring{$\alpha=3$}{Alpha=3} branes}

In the IIB theory there is one brane with tension proportional to $g_s^{-3}$, namely the 7-brane that is {the} S-dual of the D$7$-brane, and that  we denote as a $7_3$-brane following the nomenclature of \cite{Obers:1998fb}. This brane couples to an 8-form potential $E_8$, transforming with respect to the gauge parameters of the potentials $C_2$ and $D_6$, and this leads to a gauge-invariant WZ term that couples to the corresponding world-volume potentials, that are the 1-form $c_1$ and its dual 5-form 	$d_5$ \cite{Bergshoeff:2011ee}.
All other BPS branes with tension $g_s^{-3}$ can be obtained by T-duality, and they are all exotic. The corresponding mixed-symmetry potentials can be derived using the universal T-duality rules of \cite{Lombardo:2016swq}, and the outcome is that  the full $\alpha=3$ T-duality family {is  given by}
\begin{align}
& E_8 \ \ E_{8,2} \ \ E_{8,4} \ \
E_{8,6} \ \ E_{8,8} \nonumber \\
&  E_{9,2,1} \ \ E_{9,4,1} \ \
E_{9,6,1}  \ \ E_{9,8,1} \label{EfieldsIIB} \\
& E_{10,2,2} \ \ E_{10,4,2} \ \
E_{10,6,2} \ \ E_{10,8,2} \ \ E_{10,10,2}  \nonumber
\end{align}
in {the} type IIB theory and {reads}
\begin{align}
& E_{8,1} \ \ E_{8,3} \ \ E_{8,5} \ \
E_{8,7}
\nonumber \\
&  E_{9,1,1} \ \ E_{9,3,1} \ \
E_{9,5,1}  \ \ E_{9,7,1} \ \ E_{9,9,1} \label{EfieldsIIA} \\
&  E_{10,3,2} \ \ E_{10,5,2} \ \
E_{10,7,2} \ \ E_{10,9,2}   \nonumber
\end{align}
in {the} type IIA theory.

From the point of view of DFT, all the above potentials are contained in the ${\rm SO}(10,10)$ representation given by an irreducible chiral tensor-spinor $E^{MN}_\alpha$, antisymmetric in the vector indices $M$ and $N$, and with $\alpha$ labelling the $512$ spinor components.\footnote{The decomposition of the tensor-spinor $E^{MN}_\alpha$ with respect to ${\rm GL}(10,\mathbb{R})$ gives not only the potentials in \eqref{EfieldsIIB} or \eqref{EfieldsIIA}, but also additional potentials that we do not list because they do not contain  components that are  connected by T-duality to components of the potential $E_8$. From a group theory viewpoint, these representations correspond to shorter weights of the tensor-spinor representation \cite{Kleinschmidt:2011vu,Bergshoeff:2013sxa}. The contribution of these potentials is also ignored in eq. \eqref{EMNincomponents}, because we will use that equation always contracted with the brane charge, that automatically projects it on the  components for which it is correct. One can also check that eq. \eqref{EMNincomponents} restricted to these components satisfies eq. \eqref{eq:3pot}.}
 The irreducibility of the representation corresponds to the gamma-tracelessness condition
\begin{equation}
\label{eq:3pot}
\Gamma_M  E^{MN} = 0\, .
\end{equation}
In \cite{Bergshoeff:2016gub} it was shown that this DFT potential is the exotic dual \cite{Boulanger:2015mka} of the DFT RR potential $C$. As for the RR potential, one can decompose the tensor-spinor $E^{MN}_\alpha$ in terms of the  10-dimensional potentials in eq. \eqref{EfieldsIIB} or \eqref{EfieldsIIA}, introducing the Clifford vacuum $|0\rangle$ which is annihilated by the gamma matrices $\Gamma_m$. To get all the space-time potentials, one has to decompose each vector component of $E^{MN}$ as in eq. \eqref{RRpotentialsinchi}, so that one gets
\begin{align}
E^{m_1 m_2} &= \epsilon^{m_1 ...m_{10}}
\sum_p E_{m_3
...m_{10}, n_1 ...n_p} \Gamma^{n_1 ...n_p} |0\rangle \nonumber
\\
 E^{m_1}{}_{q } & = \epsilon^{m_1 ...m_{10}}
\sum_p
E_{m_2 ...m_{10}, n_1 ...n_p, q} \Gamma^{n_1 ...n_p} |0\rangle \label{EMNincomponents}
\\
E_{q_1 q_2 }  &= \epsilon^{m_1 ...m_{10}}
 \sum_p E_{m_1 ...m_{10}, n_1 ...n_p, q_1 q_2}
\Gamma^{n_1 ...n_p} |0\rangle \quad .\nonumber
\end{align}
As in the  case of the RR potentials, the chirality of the potential is fixed, and the chirality of the Clifford vacuum is  the same as the potential in {the} IIB {case} and opposite in {the} IIA {case}.

To get the WZ term for the branes charged under this potential in DFT, we first have to determine its gauge transformation with respect to the gauge parameters and field strengths of the $\alpha=1$ and $\alpha=2$ potentials. We will do this schematically to explain how the analysis of the previous section can be performed in this case as well. We write the gauge transformation with respect to the gauge parameters $\Xi^{MNPQR} $  and $\Xi^{MNP}$ of $D^{MNPQ}$ and the gauge parameter $\lambda$ of $C$ as
\begin{equation}
\delta E^{MN} = ( \Xi \cdot \Gamma )^{MN} G + (  H \cdot \Gamma )^{MN} \lambda \quad , \label{sketchygaugetransfEMN}
\end{equation}
where the products schematically denote all possible contractions that give the right index structure and that are gamma-traceless. {These}  transformations should in principle be such that the field strength
\begin{equation}
K^M = \partial_N E^{MN} + (D \cdot \Gamma )^M G + (H \cdot \Gamma )^M C \label{sketchyfieldstrengthEMN}
\end{equation}
is gauge invariant, where again the expressions are schematic. What one finds is that actually one can only impose gauge invariance for the IIB component $E_8$, while for all the other components in IIB and all the components in  IIA the section condition is not enough to make the variation of the field strength vanish. This can be understood by looking at the index structure in eq. \eqref{EMNincomponents}: one gets terms with non-vanishing coefficient containing derivatives with respect to the indices $n$ and $q$, and clearly only in the case in which none of these indices is present, which is the case of $E_8$, one gets gauge invariance. Otherwise one has to impose that these indices correspond to isometry directions.

Following the same reasoning as in the previous section, one can write down a gauge invariant WZ term for $E_8$ in DFT. In this case the world-volume is eight-dimensional, and the brane breaks ${\rm O}(10,10)$ to ${\rm O}(8,8) \times {\rm O}(2,2)$. Denoting with $a,b,... =0,....,7$ the world-volume directions, one introduces a world-volume potential $d^{abcde}$ with five indices, that transforms as a shift with respect to the pull-back on the world-volume of the gauge parameter of $D^{MNPQ}$. We will not explicitly determine the terms containing the world-volume potentials, and we schematically write  the WZ term of the  {S-dual of the D7-brane} as
\begin{equation}
S_{WZ}^{7_3} = \int d^{8}\xi \  \overline{Q}_{\hat{M} \hat{N}} [ E^{\hat{M} \hat{N}}+ ...  ]    \quad . \label{SD7braneWZterm}
\end{equation}
The charge ${Q}_{\hat{M} \hat{N}}$ is a tensor-spinor, with the vector indices antisymmetric and along the transverse directions. As in the case {of} the NS5-brane, although this expression is formally ${\rm O}(2,2)$ covariant, it is only gauge invariant for the charge ${Q}_{\hat{m}\hat{n}}$ that projects on the component $E_8$ of $E^{MN}$ with all the eight indices along the world-volume. This is the charge of the $7_3$-brane.

We can now analyse what happens if one performs a T-duality transformation.
If one performs a T-duality along a world-volume direction, say the direction $a$, the vector indices of the charge are not modified, while the spinor part changes as in the RR case, resulting in the new charge ${Q}^\prime_{\hat{m}\hat{n}} =  \Gamma^a {Q}_{\hat{m}\hat{n}}$. This corresponds to the IIA brane charged under the potential $E_{8,1}$, which is the $6_3^{(0,1)}$. The $a$ direction is an isometry direction. If{, instead, one} performs a T-duality transformation along a transverse direction, say {the} direction $\hat{n}$, then one has to consider both the action of T-duality on the vector and spinor indices, resulting in the charge $ \Gamma^{\hat{n}}{Q}_{\hat{m}}{}^{\hat{n}}$. From eq.~\eqref{EMNincomponents} one deduces that this corresponds to the IIA potential $E_{9,1,1}$, and the brane is $7_3^{({1},0)}$.

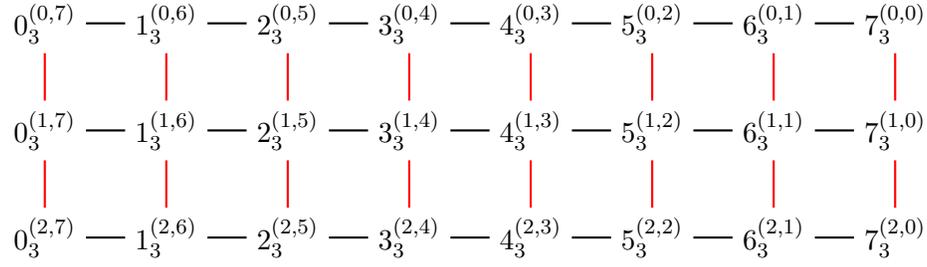
\begin{figure}[h!]
	\centering
	\begin{tikzpicture}[
	scale=0.95]
	

    \foreach \y in {0,...,7}
        \pgfmathsetmacro{\tmp}{7-\y}
        \draw (1.7*\y+2,-1) node (E\y) {$\y_3^{(0,\pgfmathprintnumber{\tmp})}$};
      \draw[thick]  (E0) -- (E1) -- (E2) -- (E3) -- (E4) -- (E5) -- (E6) -- (E7) ;	

    \foreach \y in {0,...,7}
        \pgfmathsetmacro{\tmp}{7-\y}
        \draw (1.7*\y+2,-2.5) node (E1\y) {$\y_3^{(1,\pgfmathprintnumber{\tmp})}$};
      \draw[thick]  (E10) -- (E11) -- (E12) -- (E13) -- (E14) -- (E15) -- (E16) -- (E17) ;	
     	    		
       \foreach \y in {0,...,7}
           \pgfmathsetmacro{\tmp}{7-\y}
           \draw (1.7*\y+2,-4) node (E2\y) {$\y_3^{(2,\pgfmathprintnumber{\tmp})}$};
         \draw[thick]  (E20) -- (E21) -- (E22) -- (E23) -- (E24) -- (E25) -- (E26) -- (E27) ;	

	\foreach \t in {0,...,7}
    \draw[thick, red] (E\t) -- (E1\t);	
	\foreach \t in {0,...,7}
    \draw[thick, red] (E1\t) -- (E2\t);	

	\end{tikzpicture}
	\caption{\label{tab:E_br}\sl  Branes with $g_s^{-3}$ with all T-dualities {that act} between them. The horizontal lines represent T-dualities which act on the branes in the D-brane-like way, while the vertical T-dualities act in the five-brane-like way. The first number in brackets in superscripts denotes the number of cubic directions and the second denotes the number of quadratic directions \cite{Obers:1998fb,deBoer:2012ma}. To make the pattern in the figure more transparent, the $7_3$-brane is denoted with $7_3^{(0,0)}$.}

\end{figure}

By iteration, one finds all the other $\alpha=3$ branes by T-duality starting from the $7_3${-brane}. This is summarised in
figure~\ref{tab:E_br}, where one moves horizontally performing T-dualities along the world volume and vertically performing T-dualities along the transverse directions.

\subsection{\texorpdfstring{$\alpha=4$}{Alpha=4} branes}

The prime example of an $\alpha=4$ brane is the S-dual of the D$9$-brane. In the nomenclature of \cite{Obers:1998fb}, one denotes this brane as a $9_4$-brane, that couples to the potential $F_{10}$. The other branes in the same T-duality orbit are $(9-n)^{(n,0)}$-branes, where $n$ is even in {the} IIB {case} and odd in {the} IIA {case. These branes} couple to the potentials
\begin{equation}
F_{10} \,, \ F_{10,2,2} \,, \ F_{10,4,4} \,, \ F_{10,6,6} \,, \ F_{10,8,8} \,, \ F_{10,10,10}  \label{FfieldIIB}
\end{equation}
in {the} type IIB {theory} and
\begin{equation}
F_{10,1,1} \, \ F_{10,3,3} \, \ F_{10,5,5} \, \ F_{10,7,7} \, \ F_{10,9,9} \label{FfieldsIIA}
\end{equation}
in {the} type IIA {theory}  \cite{Bergshoeff:2012ex}.

The potentials in eqs. \eqref{FfieldIIB} and \eqref{FfieldsIIA} combine in the ${\rm SO}(10,10)$ field $F_{M_1\ldots M_{10}}$, satisfying a self-duality condition. It is more useful to think of this self-dual ten-form as a symmetric irreducible bi-spinor $F_{\alpha\beta}$. Using again fermionic Fock-space notation we can then write\footnote{This is not a complete parametrisation of the bi-spinor $F$ but it contains the potentials that are related to the S-dual of the D9-brane.}
\begin{equation}
F = \epsilon^{m_1 ...m_{10}}  \sum_p  F_{m_1
...m_{10}, n_1 ...n_p, q_1 ...q_p} \Gamma^{n_1 ...n_p} \otimes \Gamma^{q_1 ...q_p} |0\rangle \otimes |0\rangle \quad , \label{FfieldDFT}
 \end{equation}
where the number of $n$ and $q$ indices is the same because of {the} irreducibility of the representation and we have used two separate chiral vacua on the right-hand side since we are dealing with a bi-spinor. If the Clifford vacuum has the same chirality {as} $F$ one gets the IIB potentials in eq. \eqref{FfieldIIB}, while if the two chiralities are opposite one gets the IIA potentials in eq. \eqref{FfieldsIIA}.

As in the $\alpha=3$ case, we want to write down the WZ term for the $9_4$-brane in DFT. We first schematically review the structure of the WZ term in {the} IIB {case}. The potential $F_{10}$ varies with respect to the parameter $\lambda_1$ of the RR 2-form potential $C_2$, and with respect to the parameter $\Xi_7$ of $E_8$.
Therefore, the WZ term contains the world-volume potentials $c_1$ and $e_7$, that transform as a shift with respect to the pull-back of the corresponding gauge parameters, and
satisfy a duality condition on the ten-dimensional world-volume. This is precisely the analysis that was performed and generalised to all dimensions in \cite{Bergshoeff:2012ex}.

One can write down the gauge transformation {of $F_{M_1\ldots M_{10}}$} in DFT in a way analogous to eq.~\eqref{sketchygaugetransfEMN}, and then find out that if one tries to construct a DFT field strength analogous to eq.~\eqref{sketchyfieldstrengthEMN}, this will only be gauge invariant for the component $F_{10}$. Again, the reason is that the gauge transformation of the putative field strength contains derivatives with respect to the indices $n$ and $q$ in eq. \eqref{FfieldDFT}, which do not vanish after imposing the strong constraint. This means that for all the mixed-symmetry potentials in eqs. \eqref{FfieldIIB} and \eqref{FfieldsIIA} one can only write down a gauge invariant WZ term after imposing that these directions are isometries. Without writing down explicitly the extra terms that make the WZ action gauge invariant, the WZ term for the $9_4$-brane in DFT is
\begin{align}
S_{WZ}^{9_4} = \int d^{10}\xi \, \overline{\overline{Q}}  [ F + ...] \,, \label{94WZtermDFT}
\end{align}
where the charge  $Q$ is a symmetric irreducible bi-spinor and the double-bar means Majorana conjugation on both spinors.  To project on the component $F_{10}$ of $F$, this charge is made of the symmetric tensor product of two Clifford vacua.

The $9_4$ brane is space-filling, so one can only perform T-dualities along world-volume directions. In particular, by T-dualising along the direction 9, the charge is rotated to
\begin{equation}
Q^\prime = \Gamma^9 \otimes \Gamma^9 Q \quad .
\end{equation}
{P}lugging this into eq. \eqref{94WZtermDFT} one finds that this projects on the IIA potential $F_{10,1,1}$, thus giving the WZ term for the $8^{(1,0)}$-brane of {the type} IIA {theory}. All the other WZ terms are also obtained by further T-dualities.

\vskip 1cm

The DFT potentials in the last two lines  of table \ref{tab:DFTpots} correspond to $\alpha=4$ mixed-symmetry potentials whose T-duality orbits of branes only contain exotic branes. The analysis performed in this section can in principle be applied to these cases as well, but one finds that there is no charge that can give a gauge-invariant WZ term if no isometries are imposed. The same applies to all the other T-duality orbits of exotic branes with higher values of $\alpha$.

\section{Conclusions}
\label{sec:concl}

In this paper we gave the explicit expressions for the WZ terms of different branes when embedded into DFT. In ordinary field theory the WZ terms of standard $(p\!-\!1)$-branes are part of effective actions that describe the dynamics of the moduli of the corresponding brane solutions in type IIA or type IIB supergravity. It would be interesting to see whether the DFT WZ terms we constructed in this paper are part of DFT effective actions  that describe the dynamics of the moduli of certain brane solutions of DFT. Some of these solutions have been investigated in the literature \cite{Berkeley:2014nza,Berman:2014jsa,Berman:2014hna} where metrics in doubled space are given. Calling our transformation of the brane as in Figure~\ref{fig:Tdual} `active', the equivalent viewpoint in those papers could be called `passive' as it changes the solution of the section constraint but keeps the brane in place.

The Wess--Zumino terms we have presented in this paper were in coordinates where the world-volume was identified directly with some of the doubled target space coordinates and thus in static gauge. Relaxing this gauge choice would require also introducing a doubled world-volume in order to have a consistent breaking of ${\rm O}(10,10)$ to ${\rm O}(p+1,p+1)\times {\rm O}(9-p,9-p)$, with an associated section constraint on the world-volume to reduce to the eventual $(p+1)$-dimensional world-volume. While writing the brane actions in such a language appears more covariant from a T-duality point of view, we have restricted in this paper to the simpler gauge-fixed formulation and leave an investigation without gauge-fixing for the future.

The effective actions for the different exotic $\alpha=2$ branes have been constructed in \cite{Kimura:2014upa} starting from the effective action of the D5-brane in IIB and performing first an S-duality transformation and then different T-dualities. A natural question to ask is whether the WZ terms that one obtains in this way coincide with the WZ terms that one obtains for various choices of the charge $Q_{\hat{M}\hat{ N}\hat{P}\hat{Q} }$ in \eqref{NS5braneWZtermgaugefixed}. We expect this not to be the case because in our formulation with manifest T-duality a crucial ingredient is that world-volume potentials and their magnetic duals are treated democratically. The duality relations \eqref{GhodgeG} between such potentials arise from introducing Lagrange multipliers that impose the Bianchi identity for a given field strength, and then solving the field equation of such field strength in terms of the field strength of the Lagrange multiplier. This implies a mixing between the kinetic term and the WZ term of the action written in terms of a single field strength. On the other hand, one expects full equivalence between the different formulations if the complete effective action is taken into account. To investigate how this is achieved, one would have to determine a manifestly T-duality covariant kinetic term for the $\alpha=2$ branes.
We leave the DFT construction of brane kinetic terms for the different values of $\alpha$ as an open project.

One of the results of this paper, {apart from the embedding of brane WZ terms into DFT},  is that we constructed the coupling of several exotic branes to a {\sl massive} IIA background. This resulted into a deformation of the results obtained  for a massless background involving the Romans mass parameter $m$. We first derived our results, using ordinary spacetime potentials, for the D-branes and the NS5-brane in the IIA theory.\,\footnote{The massive coupling of the D2-branes was given in \cite{Bergshoeff:1996cy}. We expect that the results for the NS5-brane, after making some field redefinitions, are equivalent to the results obtained earlier in the literature \cite{Bergshoeff:1997ak}.}  Next, upon making an approriate field redefinition, we embedded these results into DFT deriving the massive DFT couplings for the $\alpha={1}$ and $\alpha={2}$ branes. We only gave schematic results for the branes with $\alpha={3}$ and $\alpha={4}$ {that could involve  a non-zero Romans mass parameter in the IIA case as well}.

It is well-known that the massive couplings in the brane WZ terms have an interpretation in terms of the anomalous  creation of branes \cite{Hanany:1996ie,Bachas:1997ui}. For the massive D0-brane, this was pointed out in \cite{Danielsson:1997wq}. The WZ term in this case is given by
\begin{equation}
\label{term1}
S_{\rm massive\  D0-brane} \sim \int \ m\, b_1\, ,
\end{equation}
where $b_1$ describes the tension of a fundamental string. As explained in \cite{Danielsson:1997wq}, the presence of this term implies that,
if a D$0$-particle crosses a D$8$-brane, characterized by the Romans mass parameter $m$,
a stretched fundamental string is created, starting from the D0-brane,  in the single direction
transverse to the D8-brane. Using the notation of \cite{Bergshoeff:1996rn} this intersecting configuration is given by\footnote{Each horizontal line indicates the 10 directions $0,1,\cdots
9$ in spacetime.  A $\times  (-)$ means that the corresponding direction
is in the worldvolume of (transverse to) the brane.}
\begin{equation}
 \begin{array}{c|c}
{\rm D0}:\ \ \          \times & -    -   -   -   -  -  - -    - \\
{\rm D8}:\ \ \          \times & \times   \times  \times  \times  \times  \times \times \times   -    \\
{\rm F1}: \ \ \       \times & - - - - - - - - \times
                         \end{array}
\end{equation}

A similar situation arises for  massive NS$5$-branes in the type IIA theory. In that case there is an additional coupling to a worldvolume 6-form $c_6$ that describes the tension of a D6-brane. The strength of this coupling is proportional to $m$ and appears in the worldvolume action as
\begin{equation}
S_{\rm massive\  NS5-brane} \sim \int
m\, c_6\, .
\end{equation}
Thus, crossing a massive NS5-brane through a D8-brane a D6-brane stretched between them is created.
The corresponding intersecting configuration can be depicted as
\begin{equation}
\label{ex2}
 \begin{array}{c|c}
{\rm NS5}:\          \times & \times   \times  \times  \times  \times  -  -   -   -    \\
{\rm D8}:\ \ \          \times & \times    \times   \times \times \times  \times  \times   \times   - \\
{\rm D6}: \ \ \       \times & \times \times \times \times \times  - - - \times
                         \end{array}
\end{equation}

By T-duality we can also obtain a process involving exotic branes from this. As an example, consider two T-dualities on the last two directions in the (NS$5$, D$8$)$\rightarrow$ D$6$ configuration above. This leads to
\begin{equation}
 \begin{array}{c|c}
5_2^2:\ \ \  \      \times & \times   \times  \times  \times  \times  -  -   \otimes  \otimes    \\
{\rm D8}:\ \ \          \times & \times    \times   \times \times \times  \times  \times   - \times    \\
{\rm D6}: \ \ \       \times & \times \times \times \times \times  - -  \times -
                         \end{array}
\end{equation}
The $\otimes$ directions denote the special isometry directions of the exotic $5_2^2$ brane. This shows that exotic branes can also naturally appear in brane creation processes, as could be expected from the DFT analysis of this paper.\footnote{Given that the $5_2^2$ branes have codimension two, global consistency implies conditions for the overall charge and tension analogous to the 7-branes in the IIB theory \cite{Greene:1989ya,Vafa:1997pm}. Consistent non-geometric models involving branes of this type have originally been constructed in 
\cite{Hellerman:2002ax}.}

Let us also comment that these brane creation processes can be characterised in terms of certain root geometries in $E_{11}$~\cite{West:2001as}. Each of the individual branes appearing in the processes above can be thought of as 1/2-BPS branes and these can be associated with single real roots of $E_{11}$~\cite{Englert:2003py,Englert:2004it,Cook:2004er}. Therefore there are two real roots $\beta_1$ and $\beta_2$ corresponding to the branes passing through each other and a third real root $\beta_3$ corresponding to the brane that is created in this process.

For instance, in the example~\eqref{ex2} these roots could be chosen as
\begin{equation}
\begin{aligned}
 \beta_1&= (1,2,3,4,5,6,6,6,4,2,2) &&({\rm NS5})\\
 \beta_2&= (1,2,3,4,5,6,7,8,5,1,4) &&({\rm D8})\\
 \beta_3&= (1,2,3,4,5,6,6,6,3,1,3) &&({\rm D6})
\end{aligned}
\end{equation}

Examining the roots for all the cases above leads to the following geometry of these three roots, described by the matrix of their inner products
\begin{equation}
\beta_i \cdot \beta_j = \begin{pmatrix}2 & -2 & 0 \\ -2 & 2 & 0 \\ 0 & 0 &2 \end{pmatrix}\,.
\end{equation}
Therefore, the first two roots form an affine $\widehat{SL(2)}$ system~\cite{Englert:2007qb} while the last root is an $SL(2)$ orthogonal to it. This geometry is not sufficient to completely characterise the brane creation system: In all known examples the root $\beta_3$ is moreover invariant under those (Weyl group) U-dualities that keep the original branes in place and this characterises $\beta_3$ uniquely. It would be interesting to understand how this configuration leads to space-time solutions of supergravity or of the $E_{11}$ equations proposed in~\cite{West:2001as,Tumanov:2016abm}.

\vskip 1cm

\section*{Acknowledgements}
Two of us (E.B. and F.R.) would like to thank the IFT in Madrid,  the organisers of the workshop ``Recent Advances in T/U-dualities and Generalized Geometries'' in Zagreb and the organisers of the workshop ``Gauge theories, supergravity and superstrings'' in Benasque while E.B. and A.K. thank the \'Ecole Normale Sup\'erieure and the Institut de Physique Th\'eorique Philippe Meyerin in Paris for hospitality at an early stage of this work. We also acknowledge discussions with T. Ort\'in, who was involved at the early stages of this work,  at the IFT and in Benasque. Part of this work was also carried out at the conference ``QFTG 2018'' in Tomsk and the workshop ``Dualities and Generalised Geometries'' in Corfu. We thank the organisers of both events for having created a stimulating atmosphere.   F.R. wishes to thank the Van Swinderen Institute of the University of Groningen for hospitality. The work of E.M. is supported by the Russian state grant Goszadanie 3.9904.2017/8.9 and by the Foundation for the Advancement of Theoretical Physics and Mathematics ``BASIS'' and in part by the program of competitive growth of Kazan Federal University.

\vskip 1.5cm

\appendix

\section{Spinors of \texorpdfstring{${\rm SO}(10,10)$}{SO(10,10)}}
\label{app:Odd}

In this appendix we briefly summarise the notations we use for spinors of ${\rm SO}(10,10)$ throughout the paper. We denote with $\alpha$ and $\dot{\alpha}$ the indices of the two chiral spinor representations ${\bf 512}_S$ and ${\bf 512}_C$.
We take the ${\rm SO}(10,10)$ gamma matrices $\Gamma^M$ in the Weyl basis,
\begin{align}
\Gamma^M = \begin{pmatrix} 0 & (\Gamma^M)_{\alpha}{}^{{\dot{\beta}}}\\ (\Gamma^M )_{\dot{\alpha}}{}^\beta &0 \end{pmatrix} \quad .
\end{align}
{They satisfy} the Clifford algebra
\begin{align}
\left\{ \Gamma^M , \Gamma^N \right\} = 2 \eta^{MN}  = 2\begin{pmatrix} 0 & \mathbb{I} \\  \mathbb{I}& 0 \end{pmatrix}\quad .\label{Clifford}
\end{align}
We also introduce the charge conjugation matrix $A$ satisfying
\begin{equation}
A^{-1} \left(\Gamma_M{}\right)^T A = \Gamma_M \label{defofchargeconjugation} \quad .
\end{equation}
This matrix is antisymmetric and has the form
\begin{equation}
A = \begin{pmatrix} A^{\alpha \beta} & 0 \\ 0 & A^{\dot{\alpha} \dot{\beta}}\end{pmatrix} \quad .
\end{equation}
We choose a Majorana basis, in which all the Gamma matrices are real, and as a consequence all the spinors {can} also taken to be real.  In {the basis we are using}, the chirality matrix is defined as
\begin{equation}
\Gamma_* = \begin{pmatrix} \mathbb{I}& 0  \\   0 & -\mathbb{I} \end{pmatrix} \quad .
\end{equation}
Splitting the fundamental ${\rm SO}(10,10)$ index of $\Gamma_M$ under ${\rm GL}(10,\mathbb{R})$ as $\Gamma^M = (\Gamma^m, \Gamma_{m})$, with $m=0,...,9$, we take these matrices to satisfy
\begin{equation}
 (\Gamma^m )^{\dagger} = \Gamma_m \quad .
 \end{equation}
 As a consequence,
the matrix $A$ can be constructed as
\begin{equation}
A = \frac{1}{2^5} ( \Gamma^0 - \Gamma_0 ) (\Gamma^1 - \Gamma_1 ) \cdots (\Gamma^{9} - \Gamma_{9} ) \quad .
\end{equation}
The matrices $A$ and $\Gamma_*$ commute, stemming from the fact that one can impose a Majorana condition on Weyl spinors. To summarise, we take all the spinors to be real and chiral. Given two generic chiral spinors $\psi$ and $\phi$,  one can construct the real bilinear
\begin{equation}
\overline{\psi} \Gamma_{{M_1} ...M_{n}} \phi = \psi^T A \Gamma_{{M_1} ...M_{n}} \phi \quad .
\end{equation}
If $\psi$ and $\phi$ have the same chirality, this is  non-zero only for even $n$, while if they have opposite chirality it is non-zero only for odd $n$. Moreover, from the antisymmetry of $A$ and eq. \eqref{defofchargeconjugation} we deduce the  Majorana-flip properties
\begin{equation}
\overline{\psi} \Gamma_{M_1 ...M_n} \phi = -\overline{\phi}\Gamma_{M_n ...M_1} \psi \, ,
\end{equation}
which is non-trivial only if the spinors have the same chirality and $n$ is even or the spinors have opposite chirality and $n$ is odd.

By looking at the Clifford algebra \eqref{Clifford}, one can see that
the Gamma matrices $\Gamma^m$  and $\Gamma_m$ are proportional to the creation and annihilation operators of a fermionic harmonic oscillator, and one can therefore construct a Majorana spinor representation by declaring the Clifford vacuum $|0\rangle$ to be annihilated by the gamma matrices $\Gamma_m$:
\begin{align}
\Gamma_m |0\rangle = 0 \quad\quad \textrm{for all $m=0,\ldots, 9$} \quad .
\end{align}
The spinor module is then  generated by the $\Gamma^m$'s acting on $|0\rangle$. To construct a chiral representation, we take the Clifford vacuum to be chiral, and we only act with an even number of creation operators to construct a spinor of the same chirality of the vacuum, or an odd number of creation operators to construct a spinor of opposite chirality. This can be summarised as follows:
\begin{align}
& \psi = \sum_{p \, \textrm{even}} \frac{1}{p!} \psi_{m_1\ldots m_p} \Gamma^{m_1 \ldots m_p} |0\rangle\,, \nonumber \nonumber  \\
&  \phi = \sum_{p \, \textrm{odd}} \frac{1}{p!} \phi_{m_1\ldots m_p} \Gamma^{m_1 \ldots m_p} |0\rangle\, ,
\end{align}
where $\psi$ and $\phi$ have same and opposite chirality with respect to the vacuum{,} respectively.

The conjugate spinor is defined from a conjugate vacuum $\langle 0|$, that is annihilated by $\Gamma^m = (\Gamma_m)^\dagger$,
\begin{align}
\langle 0| \Gamma^m =0 \, ,
\end{align}
as
\begin{equation}
\overline{\psi} = \psi^T A = \langle 0| A \sum_{p } \frac{1}{p!} \psi_{m_1\ldots m_p} \Gamma^{m_p m_{p-1} \ldots m_1}
\end{equation}
where again the sum is  either {over} $p$ even or {over $p$} odd.
We normalise the vacuum such that
\begin{equation}
\langle 0 | 0 \rangle=1 \quad .
\end{equation}

\section{Gauge transformation of \texorpdfstring{$D^{MNKL}$}{D\^{}\{MNKL\}}}
\label{app:gtrm}

In this appendix we explicitly show that the gauge transformations \eqref{D6gaugetransfIIA} in {the type} IIA {theory} and \eqref{D6gaugetransfIIB} in {the type} IIB {theory} follow from {eq.~}\eqref{DMNPQgaugetransffull}, where the two different options arise from the different choice of the chirality of the Clifford vacuum.
We first consider the case of the {t}ype IIA theory, in which both the field strengths $G_{p+1}$ and the gauge parameters  $\lambda_{p-1}$ are even forms, and thus can be written as
\begin{equation}
\begin{aligned}
G&=\sum_{p=1}^3\fr{1}{p!} G_{m_1\dots m_{2p}}\Gamma^{m_1\ldots m_{2p}}|0\rangle,\\
\lambda&=\sum_{p=0}^{2}\fr{1}{p!}\lambda_{m_1\dots m_{2p}}\Gamma^{m_1\ldots m_{2p}}|0\rangle.
\end{aligned}\label{GandlambdaIIAappendix}
\end{equation}
{The}  sum has been truncated to include at most 6-forms because this is the highest rank that can occur in eq. \eqref{DMNPQgaugetransffull}. Plugging {these expressions}  in{to} \eqref{DMNPQgaugetransffull}, one gets that the term $\overline{G} \Gamma^{mnpq} \lambda$ in the  gauge trasformation of the component $D^{mnpq}$ of the potential $D^{MNPQ}$ is {given by}
\begin{align}
&\overline{G}\Gamma^{mnpq}\lambda=\\
=&\ \langle 0| \sum_{p,q=1}^{3}\fr{(-1)^{p(2p-1)}}{2^5(2p)!(2q-2)!} \Gamma_0 \cdots \Gamma_{9}\ \Gamma^{m_1\dots m_{2p}}\ \Gamma^{mnpq}\ \Gamma^{n_1\dots n_{2q-2}}|0\rangle G_{m_1\dots m_{2p}}\lambda_{n_1\dots n_{2q-2}} \, .\nonumber
\end{align}
The only contributions come from $p+q =3$, that is
\begin{align}
& \overline{G}\Gamma^{mnpq}\lambda=  \langle 0| \sum_{p	=1}^{3}\fr{(-1)^{p(2p-1)}}{2^5(2p)!(6-2p)!} \Gamma_0 \dots \Gamma_{9}\ \Gamma^{mnpq m_1\dots m_{6}}\ |0\rangle G_{m_1\dots m_{2p}}\lambda_{m_{p+1}\dots m_6} \nonumber \\
=&\ \fr{2^5}{6!}\epsilon^{mnpqm_1\dots m_6}\Big(G_{m_1\dots m_6}\lambda-\fr{6!}{4!2!}G_{m_1\dots m_4}\lambda_{m_5m_6}+\fr{6!}{2!4!}G_{m_1m_2}\lambda_{m_3\dots m_6}\Big).\label{gaugeDMNPQappendix}
\end{align}

To show that this reproduces {the transformation} \eqref{D6gaugetransfIIA} up to an overall constant, we recall that a differential  $p$-form $\omega^{(p)}$ is defined as
\begin{equation}
\omega^{(p)}=\fr{1}{p!}\omega_{m_1\dots m_p}dx^{m_1}\wedge \dots \wedge dx^{m_p} \, .
\end{equation}
The wedge product of a $p$-form and a $q$-form is defined as
\begin{equation}
\omega^{(p)}\wedge \omega^{(q)}=\fr{1}{p!q!}\omega^{(p)}{}_{m_1\dots m_p} \omega^{(q)}{}_{n_1\dots n_p}dx^{m_1}\wedge \dots \wedge dx^{n_q} \, .
\end{equation}
The components of the product of such forms then {read}
\begin{equation}
\big(\omega^{(p)}\wedge \omega^{(q)}\big)_{m_1\dots n_q}=\fr{(p+q)!}{p!q!} \omega^{(p)}{}_{[m_1\dots m_p} \omega^{(q)}{}_{n_1\dots n_p]} \, .
\end{equation}
From this,  by contracting with an epsilon symbol{,} it follows that eq. \eqref{gaugeDMNPQappendix} coincides with eq. \eqref{D6gaugetransfIIA}.

The same analysis can be repeated for the {t}ype IIB case. A T-duality transformation changes the chirality of the Clifford vacuum, and mantaining the same chirality for $G$ and $\lambda$ implies that the forms $G_{p+1}$ and $\lambda_{p-1}$ in \eqref{GandlambdaIIAappendix} now have odd rank.

To conclude this appendix, we briefly discuss the gauge transformation of the dual-graviton potential $D^{mnp}{}_q$ that arises from
T-duality along $q$. We consider this in the {type} IIA theory.
One writes
\begin{align}&
\bar{G}\Gamma^{mnp}{}_{q}\lambda=\\
=&\ \langle 0| \sum_{p,q=1}^{3}\fr{(-1)^{p(2p-1)}}{2^5 (2p)!(2q-2)!} \Gamma_0 \cdots \Gamma_{9}\ \Gamma^{m_1\dots m_{2p}}\ \Gamma^{mnp}{}_{q}\ \Gamma^{n_1\dots n_{2q-2}}|0\rangle G_{m_1\dots m_{2p}}\lambda_{n_1\dots n_{2q-2}}.\nonumber
\end{align}
There are two subtleties one faces when writing the explicit form of the above expression in terms of $p$-forms in 10 dimensions. The first one is the non-zero trace part $\Gamma^{mnp}{}_{p}=\Gamma^{mn}$, which however will vanish upon contracting with the KK5 charge $Q_{mnp}{}^q$ in the Wess-Zumino term. Since the KK5 monopole must be encoded by the same amount of degree of freedom as the smeared NS5-brane{, the}  trace part of its charge should vanish, \textit{i.e.}, $Q_{mnp}{}^p=0$. This allow{s one} to replace {$\Gamma^{mnp}{}_q$ by}  $\Gamma^{mnp}\Gamma_{q}$ and always drop the trace part from any further expressions. The same is true for {the} other non-standard branes of the T-duality orbit. The second subtlety comes from the fact that the generalized field strength $H_{MNP}$ is gauge invariant for the KK5-monopole only when an isometry direction is present, which we will always assume in what follows. The isometry direction will be chosen to be along $x^9$. Taking all this into account we obtain that the $\lambda$-dependent part of the gauge transformation of $D^{mnp}{}_9$ {reads}
\begin{equation}
\delta  D^{mnp}{}_9=\sum_{p=1}^{3}\fr{(-1)^{p(2p-1)}}{4(2p-1)!(7-2p)!} \epsilon^{m_1\dots m_{2p-1}n_1\dots n_{7-2p}}G_{m_1\dots m_{2p-1}9}\lambda_{n_1\dots n_{7-2p}9},
\end{equation}
where none of the small Latin indices is allowed to take {the} value $m=9$. {The} Hodge dual of the above tensor is the mixed symmetry potential $D_{(7,1)}$ with components defined as
\begin{equation}
\delta D_{m_1 \dots m_69,9}=\epsilon_{m_1\dots m_6 mnp9}\,\delta D^{mnp}{}_9.
\end{equation}
{Using} non-coordinate {notation} this can be written as a 6-form with two additional indices as {follows}
\begin{equation}
\delta D^{(6)}{}_{9,9}=-3 \iota_9 G^{(2)}\wedge \iota_9 \lambda^{(6)}+\iota_9 G^{(4)}\wedge \iota_9 \lambda^{(4)}-3\iota_9 G^{(6)}\wedge \iota_9 \lambda^{(2)}.
\end{equation}
Similarly{,} one can write {the} transformations of {the} gauge potentials {that couple to} the $5_2^2$ and $5_2^3$ branes {thereby} introducing 2 and 3 isometry directions{,} respectively. The first term  of the gauge transformation of the potentials{, given by  $\partial^M \Xi_{MNPQ}$,}  always gives {the}  deRahm differential of the corresponding gauge parameter upon Hodge dualization.

\section{Romans mass in DFT and the \texorpdfstring{$\alpha=2$}{alpha=2} potential}
\label{app:romans}

In this appendix we show that the ansatz of \cite{Hohm:2011cp} given in eq. \eqref{hohmkwakansatz} can be extended to the DFT potentials $D^{MNPQ}$ and $D^{MN}$ to get  the field strength $H^{mnp}$ in the presence of the Romans mass parameter. The RR DFT potential $\chi$ introduced in \cite{Hohm:2011zr} and used in \cite{Hohm:2011cp} is related to the DFT potential $C$
defined in this paper by
\begin{equation}
\chi = S_B^{-1} C \quad .\label{fieldredefusHK}
\end{equation}
In terms of $\chi$, the field strength $G$ reads
\begin{equation}
G = S_B {\partial\mkern-10mu/}\chi \label{fieldstregnthGHK}
\end{equation}
and the gauge transformation of $\chi$ with respect to the RR parameter $\lambda$ is {given by}
\begin{equation}
\delta \chi = {\partial\mkern-10mu/} \lambda \quad .\label{gaugetransfchiHK}
\end{equation}
The gauge transformation and field strength of $D^{MNPQ}$ given in eqs. \eqref{DMNPQgaugetransffull} and \eqref{HMNPDFTfieldstrength} become in terms of the potential $\chi$ (up to a redefinition of the gauge parameter $\Xi^{MNPQR}$)
\begin{equation}
\delta D^{MNPQ} = \partial_R \Xi^{RMNPQ} - \overline{\chi} \Gamma^{MNPQ} {\partial\mkern-10mu/}  \lambda \label{gaugetrasfDMNPQHKbasis}
\end{equation}
and\,\footnote{As discussed {in} section 3, we should remember that only the component $H^{mnp}$ is gauge invariant.}
\begin{equation}
H^{MNP} = \partial_Q D^{QMNP} + \overline{\chi} \Gamma^{MNP} {\partial\mkern-10mu/} \chi \quad ,\label{HMNQHKbasis}
\end{equation}
{respectively.} One can derive what the field redefinition \eqref{fieldredefusHK} means in terms of the 10-dimensional $p$-form potentials. This can be simply obtained by plugging {the mode expansion into} \eqref{fieldstregnthGHK} and \eqref{gaugetransfchiHK}{, thereby obtaining}
\begin{equation}
\delta \chi = d \lambda + d \Sigma \ \chi \qquad G = e^{-B_2} \ d \chi \quad ,
\end{equation}
where now $\chi$ denotes any redefined $p$-form RR potential.
Similarly, eqs. \eqref{gaugetrasfDMNPQHKbasis} and \eqref{HMNQHKbasis} in the IIA case lead to {the transformation rule}
\begin{equation}
\delta D_6 = d \Xi_5 - d \lambda_0 \ \chi_5 + d \lambda_2 \ \chi_3 - d \lambda_4 \ \chi_1
\end{equation}
and the gauge invariant field  strength
\begin{equation}
 H_7 = d D_6 - d \chi_1 \ \chi_5 + d \chi_3 \ \chi_3 - d \chi_5 \ \chi_1 \quad .
 \end{equation}
We are only interested in the IIA case because we want to generalise these expressions to the case $m\neq 0$, but the analogous expansion in the IIB case is obvious.

In {the basis of \cite{Hohm:2011cp}}, the Romans mass is introduced by means of the generalised Scherk-Schwarz ansatz \cite{Hohm:2011cp}
\begin{equation}
\hat{\chi} (x, \tilde{x} )=
 \chi(x) + \frac{m}{2} \tilde{x}_1 \Gamma^1  | 0\rangle \, , \label{hohmkwakansatzappendix}
\end{equation}
{A}s a result all field strengths are modified {as follows:}
\begin{equation}
G = S_B {\partial\mkern-10mu/}{\hat{\chi}}=  S_B ( {\partial\mkern-10mu/}{\chi}  +m | 0\rangle  ) \quad .
\end{equation}
We want to make a similar ansatz for $D^{MNPQ}$, that is we want to add a term linearly dependent on $\tilde{x}_1$, in such a way that implementing this ansatz together {with} \eqref{hohmkwakansatzappendix} we get a field strength $H^{mnp}$ independent {of} $\tilde{x}_1$. This actually can only be {achieved} if one also includes a linear dependence on $\tilde{x}_1$ for the potential $D^{MN}${. Indeed,} by looking at eq.~\eqref{fieldstrengthsofDsDFT} one can notice that once such violation of the section condition is allowed, this term contributes to $H^{mnp}$.

The final outcome  is that one {should use} the ansatz
\begin{equation}
\begin{aligned}
 \hat{D}^{MNPQ} (x,\tilde{x})&= D^{MNPQ} (x) + \frac{m}{2}  \tilde{x}_1 \langle 0| A\ \Gamma^1 \Gamma^{MNPQ} \chi \,, \\
 \hat{D}^{MN} (x,\tilde{x}) &= \frac{m}{2} \tilde{x}_1 \langle 0| A\ \Gamma^1 \Gamma^{MN} \chi \, .\label{SSansatzDMNPQDMN}
\end{aligned}
\end{equation}
{P}lugging this ansatz together with \eqref{hohmkwakansatzappendix} {into} the field strength
\begin{equation}
H^{MNP} = \partial_Q \hat{D}^{QMNP} + 3 \partial^{[M} \hat{D}^{NP]}  + \overline{\hat{\chi}} \Gamma^{MNP} {\partial\mkern-10mu/} \hat{\chi}
\end{equation}
one finds that the component $H^{mnp}$ does not depend on $\tilde{x}_1$ and in particular is derived from
\begin{equation}
H^{MNP} = \partial_Q D^{QMNP} + \overline{{\chi}} \Gamma^{MNP} {\partial\mkern-10mu/} {\chi} + 2 m \overline{\chi} \Gamma^{MNP} | 0\rangle \quad . \label{finalHromansappendix}
\end{equation}

We can now use the analysis of the previous appendix to expand \eqref{finalHromansappendix} in terms of the 10-dimensional RR potential. The result is
\begin{equation}
 H_7 = d D_6 - d \chi_1 \ \chi_5 + d \chi_3 \ \chi_3 - d \chi_5 \ \chi_1 + 2 m \chi_7 \quad .
 \end{equation}
This field strength is gauge invariant with respect to the gauge transformations
\begin{align}
\delta D_6 & = d \Xi_5 - d \lambda_0 \ \chi_5 + d \lambda_2 \ C_3 - d \lambda_4 \ \chi_1 + m \chi_5 \ \Sigma_1 - 2 m \lambda_6\,,
\nonumber \\
 \delta \chi &= d \lambda+ d \Sigma \ \chi + m \Sigma \quad .
\end{align}
Finally, if one rewrites these expressions in terms of the RR potentials $C$ used in this paper, one recovers eqs. \eqref{H7Romans} and \eqref{deltaD6Romans}.

\bibliographystyle{JHEP}
\bibliography{bib}

\end{document}